\documentclass[final,3p,times]{elsarticle} 

\usepackage{amsmath}
\usepackage{amssymb}
\usepackage{amsthm}
\usepackage{bm}
\usepackage{booktabs}
\usepackage{dcolumn}
\usepackage{float}
\usepackage[T1]{fontenc}
\usepackage{graphicx}
\usepackage{hyperref}
    \hypersetup{colorlinks=true,bookmarksnumbered=true}
\usepackage{makecell}
\usepackage{multicol}
\usepackage{xcolor}

\biboptions{numbers,sort&compress}   

\journal{Materials Today Physics}

\begin{document}

\begin{frontmatter}

\title{First-principles study on the \texorpdfstring{high-$T_\text{c}$}{high-Tc} superconductivity of Mg-Ti-H ternary hydrides up to the liquid-nitrogen temperature range under high pressures}

\author[affili1]{Min Pan}
\author[affili1,affili2]{Yujie Wang} 
\author[affili3]{Kaige Hu\texorpdfstring{\corref{cor1}}{}} 
\ead{hukaige@gdut.edu.cn}
\author[affili4]{Huiqiu Deng}
\texorpdfstring{\cortext[cor1]{Corresponding author.}}{}

\affiliation[affili1]{
            organization={School of Electrical Engineering, Southwest Jiaotong University},
            city={Chengdu},
            postcode={610031}, 
            country={China}}
\affiliation[affili2]{
            organization={School of Materials Science and Engineering, Southwest Jiaotong University},
            city={Chengdu},
            postcode={610031}, 
            country={China}}
\affiliation[affili3]{
            organization={School of Physics and Optoelectronic Engineering \& Guangdong Provincial Key Laboratory of Sensing Physics and System Integration Applications, Guangdong University of Technology},
            city={Guangzhou},
            postcode={510006}, 
            country={China}}
\affiliation[affili4]{
            organization={School of Physics and Electronics, Hunan University},
            city={Changsha},
            postcode={410082}, 
            country={China}}

\begin{abstract}
Ternary hydrides have emerged as the primary focus of the new wave of research into superconducting hydrides. In this work, Mg-Ti-H ternary hydrides are explored under high pressures up to 300~GPa using the prediction method of the particle swarm optimization algorithm combined with first-principles calculations. Two new structures, $P4/nmm$-MgTiH$_6$ and $Pmm2$-Mg$_3$TiH$_6$, are identified to be thermodynamically stable at both 200~GPa and 300~GPa. Thermodynamically stable structures of Mg$_3$TiH$_{12}$ are also identified, whose space groups are $R3/m$ at 200~GPa and $Pm\bar{3}m$ at 300~GPa, respectively. Among these Mg-Ti-H structures, $P4/nmm$-MgTiH$_6$ achieves a record-high $T_\text{c}$ of 81.9~K at 170~GPa, exceeding the boiling point of liquid nitrogen. Such a high $T_\text{c}$ is primarily attributed to strong electron-phonon coupling (EPC) driven by low-frequency acoustic phonon modes, with the EPC strength reaching a large value of 1.54. The $T_\text{c}$ of $Pm\bar{3}m$-Mg$_3$TiH$_{12}$ is predicted to be 40~K at 300~GPa. Furthermore, element substitution of Zr(Hf) for Ti achieves considerable enhancement of superconducting properties in our predicted hydrogen-rich and high-symmetric crystal structures, i.e., $P4/nmm$-MgTiH$_6$ and $Pm\bar{3}m$-Mg$_3$TiH$_{12}$. The high pressure required for dynamical stability is lowered to 100~GPa in both $Pm\bar{3}m$-Mg$_3$ZrH$_{12}$ and $Pm\bar{3}m$-Mg$_3$HfH$_{12}$, and to 90~GPa and 120~GPa for $P4/nmm$-MgZrH$_6$ and $P4/nmm$-MgHfH$_6$, respectively. Particularly, the electronic structure near the Fermi level is significantly modified in the $P4/nmm$-MgHfH$_6$ phase, and pronounced softening of low-frequency acoustic phonon modes occurs. As a result, the EPC strength is enhanced to 1.72, leading to a higher $T_\text{c}$ of 86~K. The element substitution strategy behaves as a promising route for discovering new high $T_\text{c}$ superconductors within known structural frameworks.
\end{abstract}


\end{frontmatter}


\section{Introduction}
\label{sec:Introduction}

The Bardeen-Cooper-Schrieffer (BCS) theory~\cite{Bardeen1957pr} and Midgal-Eliashberg theory~\cite{Eliashberg1960spj, Migdal1958spj} provide a theoretical foundation for the pursuit of high superconducting critical temperatures ($T_\text{c}$) in hydrides. Driven by both methodological advancements and groundbreaking discoveries, the present century marks a peak in hydride superconductor research. In 2014, Duan et al.~\cite{Duan2014sr} predicted that $Im\bar{3}m$-H$_3$S could exhibit a $T_\text{c}$ of 191$\sim$204~K at 200~GPa. This prediction was experimentally confirmed in the following year by Drozdov et al. ~\cite{Drozdov2015nature}, who observed superconductivity in compressed hydrogen sulfide with a $T_\text{c}$ of 203 K at 155 GPa. In 2017, Peng et al. ~\cite{Peng2017PRL} and Liu et al. ~\cite{Liu2017Julpnas} independently predicted the high-temperature superconductor $Fm\bar{3}m$-LaH$_{10}$, reporting $T_\text{c}$ of 288 K at 200 GPa and 286 K at 210 GPa, respectively. Subsequently, Somayazulu et al. ~\cite{Somayazulu2019PRL} and Drozdovet et al. ~\cite{Drozdov2019nature} successfully synthesized the predicted $Fm\bar{3}m$-LaH$_{10}$ at 190 GPa and 170 GPa with $T_\text{c}$ of 260 K and 250 K, respectively. The measurements on LaH$_{10}$ break the $T_\text{c}$ record held by H$_3$S and reach a temperature only lower than room temperature by only $\sim$40 K. These breakthroughs set a new benchmark for conventional BCS-type superconductors, and also reveal a major milestone in the field of superconductivity. More critically, the discoveries of hydrides such as H$_3$S and LaH$_{10}$ signal the emergence of a new paradigm in superconductor research, one that differs fundamentally from the traditional empirical discovery path. This paradigm emphasizes a tight interplay between theoretical predictions and experimental synthesis, enabling a more directed and efficient search for high-$T_\text{c}$ materials. There are plenty of examples of predicted hydride superconductors ~\cite{Peng2017PRL, Wang2012pnas, Li2015sr, Zhang2022prl} that have eventually been synthesized and measured, including CaH$_6$ ($T_\text{c}=215$ K at 172~GPa)~\cite{Ma2022prl}, YH$_6$ ($T_\text{c}=220$ K at 183 GPa), YH$_9$ ($T_\text{c}=243$ K at 201 GPa) ~\cite{Kong2021nc}, and LaBeH$_8$ ($T_\text{c}=110$ K at 80~GPa)~\cite{Song2023prl}, etc. It has inspired a growing wave of both theoretical and experimental efforts aiming at uncovering more high-$T_\text{c}$ hydride superconductors with near or even up to room-temperature superconductivity ~\cite{Duan2017nsr, Gorkow2018rmp, Flore2020pr, Pickard2020arcmp, Zhao2024nsr}.

As part of the effort to identify high-$T_\text{c}$ ternary hydrides, the Mg-X-H system (X denotes a second metal element other than Mg) has emerged as a promising class of high-temperature superconductors. The interest in this system is partially motivated by the systematic exploration of the binary Mg-H system ~\cite{Lonie2013prb, Feng2015rsc}, where MgH$_6$ exhibits a $T_\text{c}$ of 260 K at 300 GPa--already close to room temperature ~\cite{Feng2015rsc}. To date, a series of theoretical predictions of superconductivity have been reported for Mg-X-H systems ~\cite{Song2022jpcc, Shutov2024mtp,  Wang2025mtp, Ma2017pccp, Sukmas2020jac, Sun2019prl, Sukmas2023ijhe, Dolui2024prl, Alam2023prb, Ma2017prb, Gao2025isc, Gao2024pc, Yang2025cms, Zhang2024mtc}, among which two especially remarkable systems appear: Mg$_{0.5}$Ca$_{0.5}$H$_6$ ~\cite{Sukmas2020jac} expected to exhibit a room-temperature superconductivity  of $T_\text{c}=290$ K at 200-400 GPa, and Li$_2$MgH$_{16}$ ~\cite{Sun2019prl} predicted to exhibit the highest $T_\text{c}$ (much higher than room temperature) thus far, that is, 473 K at 250 GPa. Encouraged by these promising findings, we further explore the superconducting potential of the Mg-X-H system in this work. We choose Mg-Ti-H as our system, i.e., Ti is adopted as the third element (X$=$Ti). There are several reasons for such a choice: (1)  Ti is a transition metal element, and transition metal hydrides are capable of forming multiple stable stoichiometries and generally exhibit lower metallization pressures than other hydrides; (2) Ti has a similar electronegativity and an atomic radius comparable to Mg, suggesting that the Mg-Ti-H system may host potential high-$T_\text{c}$ superconducting phases ~\cite{Zhao2024nsr}; (3) Theoretical studies have shown that Ti-H compounds exhibit multiple stable stoichiometries ~\cite{Zhang2020prb}, while diamond anvil cell experiments have demonstrated that TiH$_2$ remains stable at room temperature up to 90 GPa ~\cite{Vennila2008ijhe, Kalita2010jap, Endo2013jac}. These considerations suggest that the Mg-Ti-H system may display a rich phase diagram and exhibit good high-$T_\text{c}$ superconductivity under high pressure.

The Mg-Ti-H system under high pressure up to 200 GPa has been investigated by Zhang et al. ~\cite{Zhang2024mtc}, where four stable hydrogen-rich phases are identified, with the highest $T_\text{c}$ predicted to be 71.2 K at 100 GPa for $R3m$-Mg$_3$TiH$_{12}$. In this work, we extend the exploration to a higher pressure range up to 300 GPa and identify three new stable phases: $P4/nmm$-MgTiH$_6$, $Pmm2$-Mg$_3$TiH$_6$, and $Pm\bar{3}m$-Mg$_3$TiH$_{12}$. Notably, $P4/nmm$-MgTiH$_6$ remains stable down to 170 GPa and exhibits a $T_\text{c}$ of 81.9 K, recording the highest $T_\text{c}$ in the Mg-Ti-H system and is the only candidate to date whose $T_\text{c}$ exceeds the boiling point of liquid nitrogen. In comparison, $Pmm2$-Mg$_3$TiH$_6$ displays a $T_\text{c}$ of 1.82 K at 200 GPa, while $Pm\bar{3}m$-Mg$_3$TiH$_{12}$ shows a $T_\text{c}$ of 40 K at 300 GPa. Furthermore, the element substitution strategy in the hydrogen-rich and high-symmetric crystal structures, i.e., $Pm\bar{3}m$-Mg$_3$TiH$_{12}$ and $P4/nmm$-MgTiH$_6$, is adopted and achieves considerable enhancement of superconducting properties. The high pressure required for dynamical stability is reduced to 100 GPa in $Pm\bar{3}m$-Mg$_3$ZrH$_{12}$ and $Pm\bar{3}m$-Mg$_3$HfH$_{12}$, and to 90 GPa and 120 GPa in $P4/nmm$-MgZrH$_6$ and $P4/nmm$-MgHfH$_6$, respectively. Particularly, owing to the significantly enhanced electron-phonon coupling (EPC) strength, the $T_\text{c}$ of $P4/nmm$-MgHfH$_6$ is predicted to be 86 K, even higher than the record predicted in the Mg-Ti-H system.

\section{Computational methods}
\label{sec:Method}

The CALYPSO (Crystal structure AnaLYsis by Particle Swarm Optimization) package ~\cite{Wang2010prb, Wang2012cpc} is adopted to search high-pressure crystal structures. Structure relaxations and total energies are determined in the framework of density functional theory with Perdew-Burke-Ernzerhof (PBE) parametrization ~\cite{Perdew1996prl} of the generalized gradient approximation (GGA) as implemented in the Vienna \textit{ab initio} simulation package (VASP) ~\cite{Kresse1996prb}. The electronic structure is calculated using DS-PAW software, which adopts the projector augmented-wave (PAW) method to describe ion-electron interactions ~\cite{Blochl1994prb}. $3s^23p^63d^24s^2$, $2p^63s^2$, and $1s^1$ electrons are treated as valence electrons for Ti, Mg, and H, respectively. The cutoff energy is set to be 550 eV. The Brillouin zone is sampled with a $\bm{k}$-point mesh of $2\pi\times 0.03$~\AA$^{-1}$.

Phonon spectrum and the EPC matrix elements are calculated with linear-response theory embedded in Quantum ESPRESSO (QE) ~\cite{Giannozzi2009jpcm, Giannozzi2017jpcm}. Norm-conserving pseudopotentials ~\cite{Hamann2013prb} for Mg, Ti, and H are selected with a kinetic energy cutoff of 80 Ry with a Methfessel-Paxton smearing of 0.02 Ry. A $24\times 24\times 24$ $\bm{k}$-point grid for electric bands calculations and a $3\times 3 \times 3$ $\bm{q}$-point grid for phonon dispersions calculations are used for $Pmm2$-Mg$_3$TiH$_6$,  a $24\times 24 \times 24$ $\bm{k}$-point grid and a $4\times 4\times 4$ $\bm{q}$-point are used for $P4/nmm$-MgTiH$_6$, $R3m$-Mg$_3$TiH$_{12}$ and $P4/nmm$-MgTiH$_{10}$, while a $24\times 24\times 24$ $\bm{k}$-point grid and a $6 \times 6 \times 6$ $\bm{q}$-point are used for $Pm\bar{3}m$-Mg$_3$TiH$_{12}$, $I4_1amd$-MgTiH$_8$, $Pm\bar{3}m$-Mg$_3$ZrH$_{12}$ and $Pm\bar{3}m$-Mg$_3$HfH$_{12}$. 
The superconducting critical temperature $T_\text{c}$ is estimated via the Allen-Dynes-modified McMillan formula ~\cite{Allen1975prb}
\begin{equation} \label{eq:Tc}
T_\text{c}=\frac{f_1f_2\omega_{\log}}{1.2}\exp\left(-\frac{1.04(1+\lambda)}{\lambda-\mu^*(1+0.62\lambda)}\right),
\end{equation}
where $\omega_{\log}$ is the logarithmic average frequency, $\lambda$ is the EPC parameter, $\mu^*$ is the effective Coulomb repulsion whose typical value ranges from 0.1 to 0.13 ~\cite{Liang2019prb, Zhao2022prb, Zhang2020prb_Structure}, and $f_1$ ($f_2$) is the strong coupling (shape correction) factor. $\omega_{\log}$ and $\lambda$ are defined as
\begin{equation}
\omega_{\log} = \exp\left(\frac{2}{\lambda}\int\frac{d\omega}{\omega}\alpha^2F(\omega)\ln\omega\right),\quad \lambda=2\int\frac{\alpha^2F(\omega)}{\omega}d\omega,
\end{equation}
where $\omega$ is the phonon frequency and $\alpha^2F(\omega)$ is the Eliashberg spectral function. $\alpha^2F(\omega)$ is described as 
\begin{equation}
\alpha^{2}F\left(\omega\right)=\frac{1}{2}\sum_{\bm{q}\nu}W_{\bm{q}}\omega_{\bm{q}\nu}\lambda_{\bm{q}\nu}\delta\left(\omega-\omega_{\bm{q}\nu}\right), \label{eq:a2f}
\end{equation}
where $\bm{q}$ is the phonon vector, $\nu$ is the index of phonon dispersions, $W_{\bm{q}}$ is the weight parameter, $\omega_{\bm{q}\nu}$ is the mode-resolved phonon frequency, and $\lambda_{\bm{q}\nu}$ is the mode-resolved EPC constant. The expression of $\lambda_{\bm{q}\nu}$ is
\begin{equation} \label{eq:lambda_qv}
\lambda_{\bm{q}\nu} = \frac{\gamma_{\bm{q}\nu}}{\pi N(\epsilon_F)\omega_{\bm{q}\nu}^2},
\end{equation}
where $N(\epsilon_F)$ denotes the density of states at the Fermi level and $\gamma_{\bm{q}\nu}$ is the phonon linewidth. $\gamma_{\bm{q}\nu}$ is expressed as
\begin{equation}
\gamma_{\bm{q}\nu} = 2\pi\omega_{\bm{q}\nu}\sum_{mn}\sum_{\bm{k}}\left | g_{mn}^{\nu}(\bm{k}, \bm{q}) \right |^2\delta(\epsilon_{\bm{k+q},m}-\epsilon_\text{F})\delta(\epsilon_{\bm{k},n}-\epsilon_\text{F}),
\end{equation}
where $m(n)$ is the index of electronic energy bands, $\bm{k}$ is the electron vector, $\epsilon$ is the energy of electron Bloch states, $\epsilon_\text{F}$ is the Fermi energy, and $g_{mn}^{\nu}(\bm{k}, \bm{q})$ represents the EPC matrix element. The expression of $g_{mn}^{\nu}(\bm{k}, \bm{q})$ is
\begin{equation}\label{eq:gmnnuk}
g_{mn}^{\nu}(\bm{k}, \bm{q}) = \sqrt{\frac{\hbar}{2\omega_{\bm{q}\nu}}}\left \langle \psi_{m\bm{k}+\bm{q}}\left | \Delta_s^{\bm{q}\nu}e^{i\bm{q}\cdot\bm{r}} \right |\psi_{n\bm{k}}\right \rangle,
\end{equation}
where $\psi$ is the electronic Kohn-Sham states, $\bm{r}$ is the electron position, and $\Delta_s^{\bm{q}\nu}e^{i\bm{q}\cdot\bm{r}}$ is the finite variation in the self consistent potential corresponding to a phonon displacement of wavevector $\bm{q}$ and mode index $\nu$ ~\cite{Flore2020pr}.
The definitions of $f_1$ and $f_2$ are
\begin{equation}
f_1=\sqrt[3]{1+\left(\frac{\lambda}{2.46(1+3.8\mu^*)}\right)^{\frac{3}{2}}}, 
\end{equation}
\begin{equation}
f_2=1+\frac{\left(\frac{\bar{\omega}_2}{\omega_{\log}}-1\right)\lambda^2}{\lambda^2+\left(1.82(1+6.3\mu^*)\frac{\bar{\omega}_2}{\omega_{\log}}\right)^2},
\end{equation}
where $\bar{\omega}_2=\sqrt{\frac{2}{\lambda}\int_0^\omega \alpha^2F(\omega)\text{d}\omega}$ is the second moment of the  normalized weight function.

The isotropic Migdal-Eliashberg equations are solved through the implementation of the Electron-Phonon Wannier (EPW) code  ~\cite{Ponce2016cpc, Lee2023npj}. The EPW code is used to perform interpolation calculations of the EPC matrix elements and then obtain the EPC constant $\lambda$. The Dirac $\delta$ functions for electrons and phonons are approximated using Gaussian functions with a width of 100 meV and 0.5 meV, respectively. The sum over the Matsubara frequencies is truncated at ten times the highest phonon frequency, and the width of the Fermi surface window is set to four times the highest phonon frequency. The perfect match between the Wannier-interpolated and DFT-calculated band structures of these structures is shown in Fig. S1 of the supplemental material (SM).

We solve the isotropic Midgal-Eliashberg equations on the imaginary axis, which takes the form of a set of nonlinear coupled equations for the mass renormalization function and the superconducting gap, with temperature included as a constant in the equations ~\cite{Margine2013prb}:
\begin{equation}
Z\left ( i\omega _{j}  \right ) = 1 + \frac{\pi T}{\omega_j}\sum_{j^{\prime}}\frac{\omega_{j^{\prime}}}{\sqrt{\omega_{j^{\prime}}^2+\Delta^2\left ( i\omega_j \right )}}\lambda\left ( \omega_j - \omega_{j^{\prime}}\right ),
\end{equation}
\begin{equation}
Z\left ( i\omega _{j}  \right )\Delta\left(i\omega_j\right) = \pi T\sum_{j^{\prime}}\frac{\Delta\left(i\omega_{j^{\prime}}\right)}{\sqrt{\omega_{j^{\prime}}^2+\Delta^2\left(i\omega_j\right)}}\left[\lambda\left(\omega_j-\omega_{j^{\prime}}\right)-\mu^*_c\right],
\end{equation}
where $T$ represents the temperature, $Z\left(\omega_j\right)$ denotes the mass renormalization function, and $\Delta\left(i\omega_j\right)$ represents the superconducting gap function. The superconducting critical temperature is defined as the temperature at which the superconducting gap vanishes.

$T_\text{c}^{\text{ML}}$ is calculated by machine learning based on Midgal-Eliashberg theory, which is defined by Xie et al. ~\cite{Xie2022npj} as:
\begin{equation}
T_\text{c}^{\text{ML}} = \frac{f_\omega f_\mu\omega_{\text{log}}}{1.20}\exp\left(-\frac{1.04(1+\lambda)}{\lambda-\mu^*(1+0.62\lambda)}\right),
\end{equation}
where $f_\omega$ and $f_\mu$ are defined as 
\begin{equation}
f_\omega = 1.92\left (\frac{\lambda + \frac{\omega_{\text{log}}}{\bar{\omega}_2}-\sqrt[3]{\mu^*}}{\sqrt{\lambda}\exp\left( \frac{\omega_{\text{log}}}{\bar{\omega}_2}\right)} \right) - 0.08,
\end{equation}
\begin{equation} 
f_\mu = \frac{6.86\exp\left( \frac{-\lambda}{\mu^*}\right)}{\frac{1}{\lambda}-\mu^*-\frac{\omega_{\text{log}}}{\bar{\omega}_2}} + 1.
\end{equation}

\section{Results}

\subsection{Predicted stable and metastable structures of the Mg-Ti-H system}

\begin{figure*}[th]
\centering
\includegraphics[width=12cm]{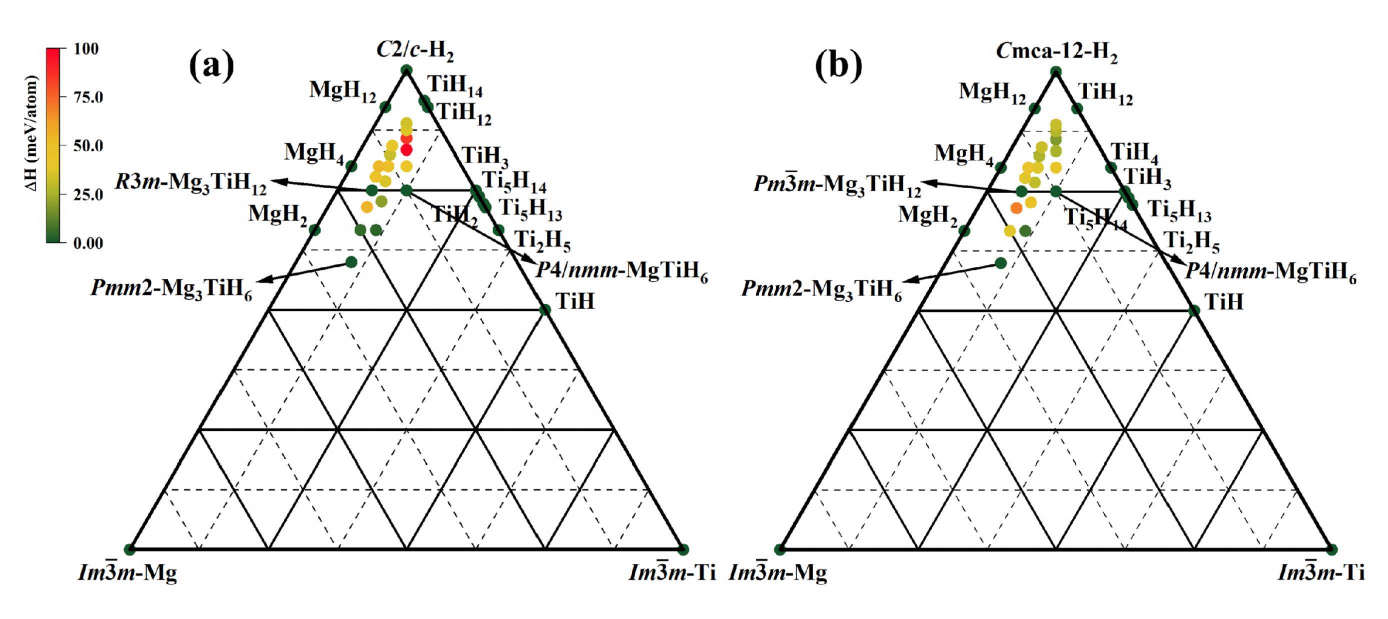}
\caption{Ternary phase diagrams (convex hull) of the Mg$_x$TiH$_{2y}$ ($x=1$-3, $y=3$-8) system at (a) 200 and (b) 300 GPa, respectively. The color of each structure represents its energy difference from the convex hull.}
\label{fig1}
\end{figure*}

The random structural searches for the Mg$_x$TiH$_{2y}$ ($x=1$-3, $y=3$-8) system are performed at 200 and 300 GPa, respectively. Fig. \ref{fig1} shows the results on the ternary phase diagrams. To ensure the thermodynamic stability of our ternary hydrides, the formation enthalpy convex hull is constructed with known thermodynamical stable structures as references, including binary systems Mg-H ~\cite{Abe2018prb, Li2017pccp}, Ti-H ~\cite{Zhang2020prb}, and element systems Mg ~\cite{Liu2009jap}, Ti ~\cite{Zhang2020prb}, and H~\cite{Pickard2007np} under the same pressures. As shown in Figs. \ref{fig1}(a) and \ref{fig1}(b), two new ternary hydrides, $P4/nmm$-MgTiH$_6$ and $Pmm2$-Mg$_3$TiH$_6$, are thermodynamically stable at both 200 and 300 GPa: They all lie on the ternary convex hull with an $E_\text{Hull}$ of 0. Further analysis confirms that $P4/nmm$-MgTiH$_6$ and $Pmm2$-Mg$_3$TiH$_6$ are thermodynamically stable in the whole range of 200$\sim$300 GPa. In Ref.~\cite{Zhang2024mtc}, the stable MgTiH$_6$ structure was reported to be $Pmmn$-MgTiH$_6$ at 200 GPa, which we also identified; however, its energy is 5 meV/atom higher than that of the $P4/nmm$-MgTiH$_6$ phase we discovered in this work. Therefore, $P4/nmm$-MgTiH$_6$ is more likely to be synthesized at 200 GPa, and we adopt this structure.\footnote{Another reason that we do not further investigate $Pmmn$-MgTiH$_6$ here is that it has already been studied in Ref. ~\cite{Zhang2024mtc} (below 200 GPa), and the predicted $T_\text{c}$ is \emph{relatively} low ($37.7\sim 45.6$ K for $\mu^*=0.1\sim0.13$).} At 200 GPa, we also identify the thermodynamically stable phase $R3m$-Mg$_3$TiH$_{12}$, which was first predicted in Ref.~\cite{Zhang2024mtc}. Since $R3m$-Mg$_3$TiH$_{12}$ possesses the highest $T_\text{c}$ (71.2 K at 100 GPa) predicted in Ref.~\cite{Zhang2024mtc}, we also examined this phase in the extended pressure region up to 300 GPa. We find that $Pm\bar{3}m$-Mg$_3$TiH$_{12}$ becomes the new thermodynamically stable phase at 300 GPa instead of $R3m$-Mg$_3$TiH$_{12}$. In addition, certain thermodynamically metastable phases may still be interesting for the research on their superconductivity, provided they are dynamically stable, since such structures are also possible to be synthesized experimentally. We identified two metastable phases, $I4_1amd$-MgTiH$_8$ and $P4/nmm$-MgTiH$_{10}$, both of which exhibit well-formed hydrogen cage structures and have $E_\text{Hull}$ values below 70 meV/atom. In the following, we will focus on six structures, i.e., four stable structures $P4/nmm$-MgTiH$_6$, $Pmm2$-Mg$_3$TiH$_6$, $R3m$-Mg$_3$TiH$_{12}$, and $Pm\bar{3}m$-Mg$_3$TiH$_{12}$, and two metastable structures $I4_1amd$-MgTiH$_8$ and $P4/nmm$-MgTiH$_{10}$. 

\begin{figure*}[th]
\centering
\includegraphics[width=10cm]{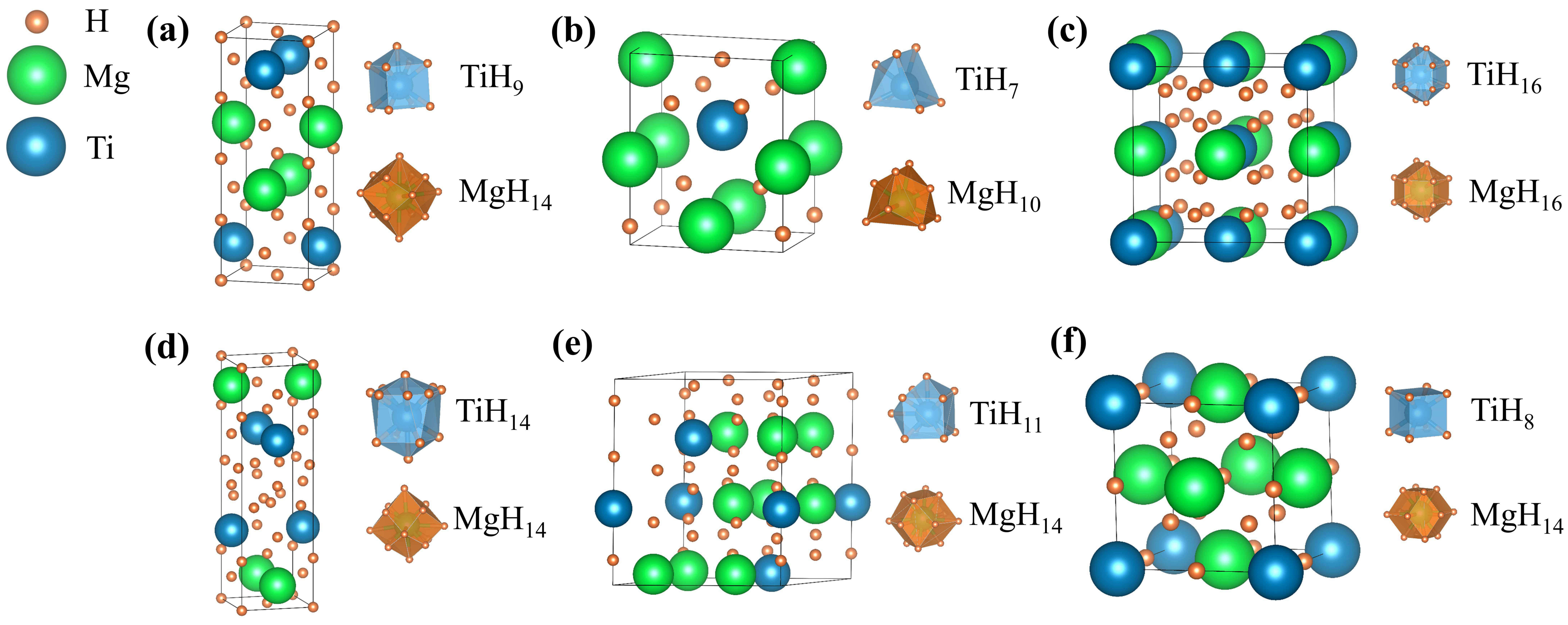}
\caption{Crystal structures of (a) $P4/nmm$-MgTiH$_6$ at 200 GPa, (b) $Pmm2$-Mg$_3$TiH$_6$ at 200 GPa, (c) $I4_1amd$-MgTiH$_8$ at 300 GPa, (d) $P4/nmm$-MgTiH$_{10}$ at 300 GPa, (e) $R3m$-Mg$_3$TiH$_{12}$ at 200 GPa, and (f) $Pm\bar{3}m$-Mg$_3$TiH$_{12}$ at 300 GPa.}
\label{fig2}
\end{figure*}

Figure \ref{fig2} illustrates the crystal structures of the six Mg-Ti-H phases. Detailed crystallographic information, including lattice constants, fractional atomic coordinates, and Wyckoff positions, is listed in Table S1 of the SM. In $P4/nmm$-MgTiH$_6$, metal atoms Mg and Ti both locate in lattice faces, forming TiH$_9$ and MgH$_{14}$ units, respectively [Fig. \ref{fig2}(a)]. In $Pmm2$-Mg$_3$TiH$_6$, the structure contains irregular TiH$_7$ and MgH$_{10}$ polyhedral units due to its low symmetry [Fig. \ref{fig2}(b)]. The metastable $I4_1amd$-MgTiH$_8$ consists of (Mg/Ti)-H$_{16}$ units, in which each H$_{16}$ cage is formed by ten rectangles and eight triangles [Fig. \ref{fig2}(c)]. In the metastable $P4/nmm$-MgTiH$_{10}$ phase, Mg and Ti locate in different H$_{14}$ cages [Fig. 2(d)]. Consistent with Ref. ~\cite{Zhang2024mtc}, the hydrogen atoms in $R3m$-Mg$_3$TiH$_{12}$ form H$_{11}$ and H$_{14}$ units around the Ti and Mg sites, respectively [Fig. \ref{fig2}(e)]. In $Pm\bar{3}m$-Mg$_3$TiH$_{12}$ [Fig. \ref{fig2}(f)], the Ti atoms are distributed on the vertices and surrounded by H$_{8}$ units composed of eight square H frames and Mg atoms are distributed in the center of the face. The H$_{14}$ units, composed of eight irregular quadrilateral H frames and four parallelogram H frames, surround the Mg atoms. Moreover, the $Pm\bar{3}m$-Mg$_3$TiH$_{12}$ phase can be considered to be obtained by replacing the Ti atom on the $Fm\bar{3}m$-TiH$_3$ face center with the Mg atom.

\subsection{Electronic properties and superconductivity of the Mg-Ti-H phases}

\begin{figure}[th]
\centering
\includegraphics[width=7cm]{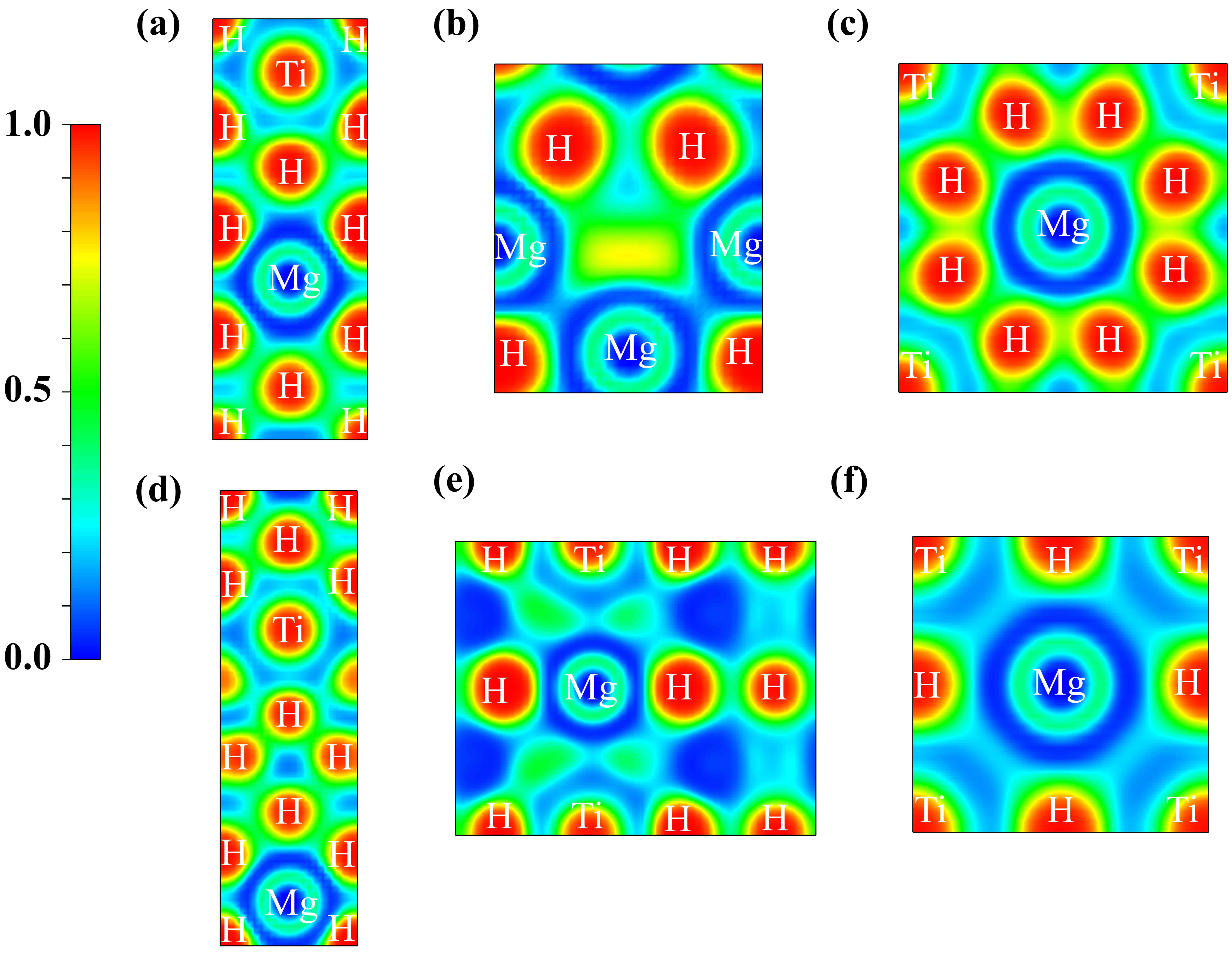}
\caption{Electron localization function (ELF) of (a) the (100) plane of $P4nmm$-MgTiH$_6$ at 200 GPa, (b) the (100) plane of $Pmm2$-Mg$_3$TiH$_6$ at 200 GPa, (c) the (001) plane of $I4_1amd$-MgTiH$_8$ at 300 GPa, (d) the (100) plane of $P4nmm$-MgTiH$_{10}$ at 300 GPa, (e) the (010) plane of $R3m$-Mg$_3$TiH$_{12}$ at 200 GPa, and (f) the (001) plane of $Pm\bar{3}m$-Mg$_3$TiH$_{12}$ at 300 GPa.}
\label{fig3}
\end{figure}

\begin{figure*}[th]
\centering
\includegraphics[width=15cm]{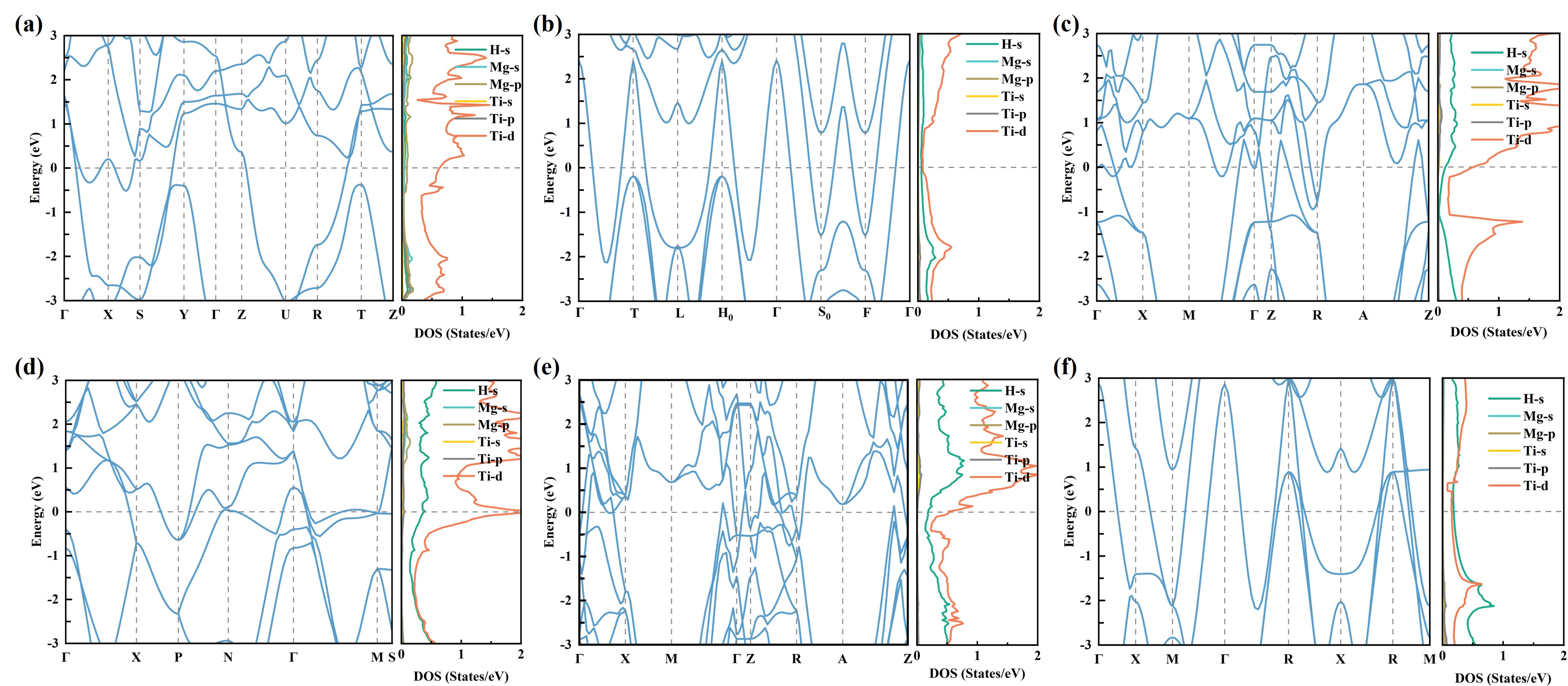}
\caption{Electronic band structure and PDOS of (a) $Pmm2$-Mg$_3$TiH$_6$ at 200 GPa, (b) $R3m$-Mg$_3$TiH$_{12}$ at 200 GPa, (c) $P4/nmm$-MgTiH$_6$ at 200 GPa, (d) $I4_1amd$-MgTiH$_8$ at 300 GPa, (e) $P4/nmm$-MgTiH$_{10}$ at 300 GPa, and (f) $Pm\bar{3}m$-Mg$_3$TiH$_{12}$ at 300 GPa. The Fermi level is set to zero.}
\label{fig4}
\end{figure*}

To analyze the chemical bonding, the electron localization function (ELF) and Bader charge are calculated. Bader's analysis shows that charge transfer from Mg and Ti to H occurs in all phases except $Pmm2$-Mg$_3$TiH$_6$, where instead each Ti atom gains $\sim0.1e$ from Mg atoms ($e$ is the electron charge, i.e., here $e<0$). Fig. \ref{fig3} illustrates that the ELF values near Mg and Ti atoms in the six compounds are close to zero, indicating delocalized electrons and the formation of ionic bonds between the metal atoms and hydrogen. $P4/nmm$-MgTiH$_6$, $Pmm2$-Mg$_3$TiH$_6$, $P4/nmm$-MgTiH$_{10}$, and $R3m$-Mg$_3$TiH$_{12}$ exhibit ELF values of $\sim$0.4 between neighboring H-H atoms, similar to those in $Pm\bar{3}m$-Mg$_3$TiH$_{12}$, suggesting the absence of electron localization and covalent bonding. In $P4/nmm$-MgTiH$_6$, $Pmm2$-Mg$_3$TiH$_6$, $P4/nmm$-MgTiH$_{10}$, $R3m$-Mg$_3$TiH$_{12}$, and $Pm\bar{3}m$-Mg$_3$TiH$_{12}$, the ELF values between H-H atoms are approximately 0.4 [Figs. \ref{fig3}(a-b, d-f)], which is characteristic of metallic bonds~\cite{Koumpourasiop2020jpcm}. In contrast, $I4_1amd$-MgTiH$_8$ shows an ELF value of $\sim$0.7 between the nearest H-H atoms, indicating the presence of weak covalent bonding [Fig.~\ref{fig3}(c)].

Figure \ref{fig4} shows the electronic band structures and projected density of states (PDOS) of the six Mg-Ti-H compounds, clearly demonstrating the metallic character of all these phases. At the Fermi level, the contribution of the electrons of Mg to the density of states (DOS) is almost negligible, the same as the behaviors reported in other Mg-Ti-H compounds~\cite{Zhang2024mtc}. For the $Pmm2$-Mg$_3$TiH$_6$ phase, the DOS at the Fermi level originates almost exclusively from $d$ electrons of Ti [Fig. \ref{fig4}(a)]; in contrast, the DOS near the Fermi level in the other phases is predominantly dominated by the $d$ electrons of Ti with $s$ electrons of H. The electronic band structure of $I4_1amd$-MgTiH$_8$ exhibits minimal dispersion along the $\Gamma$-M-S direction, forming a flat region that results in a Van Hove singularity in the DOS [Fig.~\ref{fig4}(d)]. Consequently, this phase displays the highest total density of states (TDOS) among the six phases. However, the contribution of the $s$-electrons of H at the Fermi level ($N_\text{F}^\text{H}$) is only 11.6 \%, and most of the contribution comes from the $d$-electron of Ti, which could be unfavorable for H-driven superconductivity. For $R3m$-Mg$_3$TiH$_{12}$ and $Pm\bar{3}m$-Mg$_3$TiH$_{12}$, the contribution from the $s$-electrons of H is comparable to that from the $d$-electrons of Ti at the Fermi level [Figs. \ref{fig4}(b, f)]. However, the TDOS at the Fermi level is low, which may be insufficient to support the formation of a large number of Cooper pairs.

\begin{figure*}[th]
\centering
\includegraphics[width=15cm]{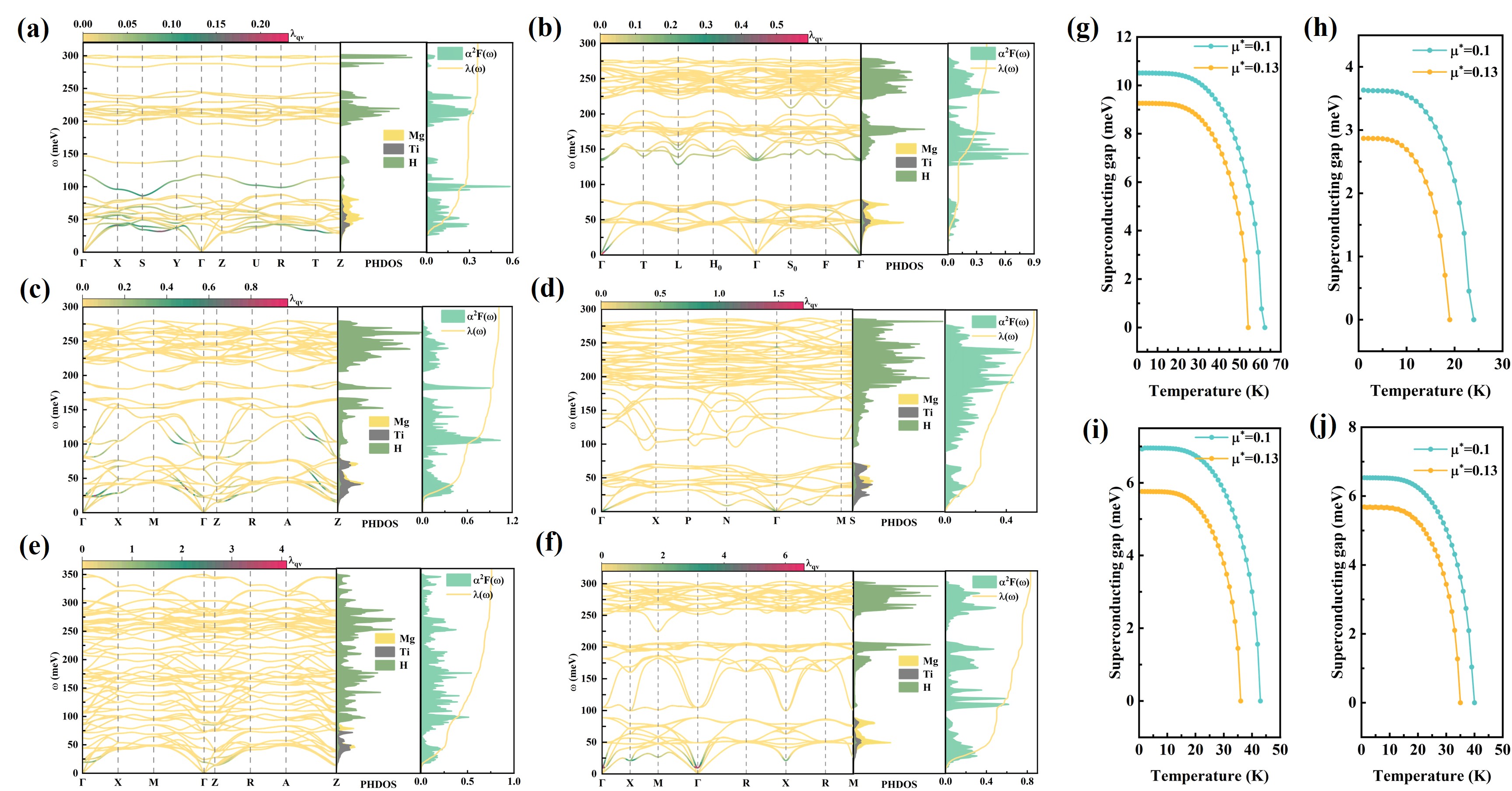}
\caption{Calculated phonon dispersion curves, projected phonon density of states, $\alpha^2F(\omega)$ and electron-phonon integrals $\lambda$ for (a) $Pmm2$-Mg$_3$TiH$_6$ at 200 GPa, (b) $R3m$-Mg$_3$TiH$_{12}$ at 200 GPa, (c) $P4/nmm$-MgTiH$_6$ at 200 GPa, (d) $I4_1amd$-MgTiH$_8$ at 300 GPa, (e) $P4/nmm$-MgTiH$_{10}$ at 300 GPa, and (f) $Pm\bar{3}m$-Mg$_3$TiH$_{12}$. The isotropic superconducting gaps at various temperatures for (g) $P4/nmm$-MgTiH$_6$ at 200 GPa, (h) $I4_1amd$-MgTiH$_8$ at 300 GPa, (i) $P4/nmm$-MgTiH$_{10}$ at 300 GPa, and (j) $Pm\bar{3}m$-Mg$_3$TiH$_{12}$ at 300 GPa. The colors mapped in the phonon dispersion curves indicate the magnitude of $\lambda_{\bm{q}\nu}$.} 
\label{fig5}
\end{figure*}

To explore possible superconductivity, we further calculate the phonon dispersion curves, projected phonon density of states (PHDOS), Eliashberg spectral function $\alpha^2F(\omega)$, with the electron-phonon integrals $\lambda$ for the six Mg-Ti-H compounds. The results are shown in Fig. \ref{fig5}. Moreover, the mode-resolved electron-phonon coupling constants $\lambda_{\bm{q}\nu}$ are calculated using Eq. \eqref{eq:lambda_qv}. The phonon dispersion curves of the six phases at their respective pressures exhibit no imaginary frequencies, indicating the dynamical stability. The low-frequency modes are mainly attributed to vibrations of Mg and Ti atoms, while hydrogen vibrations dominate the high-frequency modes. For the $Pmm2$-Mg$_3$TiH$_6$ and $R3m$-Mg$_3$TiH$_{12}$ phases, the EPC strength is relatively weak, with a $\lambda$ value of 0.36 and 0.40, respectively [Figs.~\ref{fig5}(a, b)]. This weakness primarily stems from the low contribution of H-1$s$ states at the Fermi level, which reduces coupling to H-related phonon modes. In the $P4/nmm$-MgTiH$_6$ phase, the H-derived vibrations in the mid- and high-frequency region ($>75$ meV) contribute approximately 45\% to $\lambda$, while the low-frequency region (0-75 meV), dominated by Mg and Ti vibrations, accounts for about 55\% [Fig.~\ref{fig5}(c)]. In addition, several soft phonon modes are observed along the M-$\Gamma$ and A-Z paths, which make significant contributions to the EPC. For the $I4_1amd$-MgTiH$_8$ phase, the EPC parameter $\lambda$ is only 0.58 [Fig.~\ref{fig5}(d)]. Although it exhibits a relatively high DOS at the Fermi level [see Fig.~\ref{fig4}(d)], those states are mainly contributed by Ti-$3d$ orbitals. The coupling between the H-dominated mid- and high-frequency phonon modes and the Ti-$3d$ electrons is weak, leading to a small $\lambda$. For the $P4/nmm$-MgTiH$_{10}$ and $Pm\bar{3}m$-Mg$_3$TiH$_{12}$ phases, the low-frequency region contributes 37\% and 59\% to $\lambda$, respectively, while the H-dominated mid- and high-frequency modes account for 63\% and 41\% [Figs.~\ref{fig5}(e, f)]. It is noteworthy that although the maximum mode-resolved electron-phonon coupling constants ($\lambda_{\bm{q}\nu}^{\text{max}}$) in $P4/nmm$-MgTiH$_{10}$ and $Pm\bar{3}m$-Mg$_3$TiH$_{12}$ are significantly enhanced, their $\lambda$ values remain lower than that of $P4/nmm$-MgTiH$_6$. This is a reflection that the EPC parameter $\lambda$ is not governed by a single strongly coupled phonon mode but rather determined by the collective coupling of all phonon modes with the electronic states.

\begin{table*}[th]
\centering
\caption{The calculated values of EPC constant $\lambda$, $\lambda_{\bm{q}\nu}^{\text{max}}$, and $T_\text{c}$. \label{tab1}}
\begin{tabular}{*{8}{c}}
\toprule[1pt]
Compound & Space group & Pressure & $\lambda$ & $\lambda_{\bm{q}\nu}^{\text{max}}$ & $T_\text{c}^{\text{ADM}}$ (K) & $T_\text{c}^{\text{ML}}$ (K) & $T_\text{c}^{\text{E}}$ (K) \\
\midrule
MgTiH$_6$ & $P4/nmm$ & 170 GPa & 1.54 & 4.83 & 61.9-69.0 & 72.6-81.9 &  71-75 \\
MgTiH$_6$ & $P4/nmm$ & 200 GPa & 1.02 & 0.98 & 47.0-54.5 & 51.3-61.2  & 54-62 \\
Mg$_3$TiH$_6$ & $Pmm2$ & 200 GPa & 0.36 & 0.23 & 0.6-1.8 & 0.6-1.6 & / \\
Mg$_3$TiH$_{12}$ & $R3m$ & 200 GPa & 0.40 & 0.59 & 2.9-6.5 & 2.9-6.3& / \\
MgTiH$_8$ & $I4_1amd$ & 300 GPa & 0.58 & 1.71 & 14.6-21.3 & 12.8-18.9 & 19-24 \\
MgTiH$_{10}$ & $P4/nmm$ & 300 GPa & 0.76 & 4.1 & 31.4-39.6 & 30.8-39.9 & 36-44 \\
Mg$_3$TiH$_{12}$ & $Pm\bar{3}m$ & 300 GPa & 0.83 & 6.62 & 31.1-38.3 & 29.5-37.4 & 35-40 \\
MgZrH$_6$ & $P4/nmm$ & 90 GPa & 1.13 & 1.07 & 50.4-57.4 & 57.4-67.3 &  62-70 \\
MgHfH$_6$ & $P4/nmm$ & 120 GPa & 1.72 & 8.02 & 50.2-54.4 & 74.3-82.6 &  78-86 \\
Mg$_3$ZrH$_{12}$ & $Pm\bar{3}m$ & 100 GPa & 0.81 & 3.12 & 32.6-40.5 & 31.4-39.6 & 39-45 \\
Mg$_3$HfH$_{12}$ & $Pm\bar{3}m$ & 100 GPa & 0.83 & 5.27 & 28.2-34.4 & 28.4-35.8 & 34-40 \\
\bottomrule[1pt]
\end{tabular}
\end{table*}

We then calculate the $T_\text{c}$s of the six Mg-Ti-H compounds. Three distinct equations are employed. Among the results, $T_\text{c}^{\text{ADM}}$ is calculated using the Allen-Dynes modified McMillan equation, $T_\text{c}^{\text{ML}}$ is predicted by a machine-learning model based on the Migdal-Eliashberg theory, and $T_\text{c}^{\text{E}}$ is obtained by self-consistently solving the isotropic Migdal-Eliashberg equation, with the Coulomb pseudopotential $\mu^*$ ranging from 0.10 to 0.13. The results are summarized in lines 2$\sim$7 of Table \ref{tab1} (other data in the table will be explained later). For the $Pmm2$-Mg$_3$TiH$_6$ and $R3m$-Mg$_3$TiH$_{12}$ phases, the $\lambda$ values are calculated to be quite small and $T_\text{c}^\text{ADM}$ and $T_\text{c}^\text{ML}$ are consequently very low. Therefore, $T_\text{c}^\text{E}$ values are also expected to be low, and we do not further calculate these values using the corresponding isotropic Migdal-Eliashberg equations, as such calculations are typically computationally intensive. For the other four compounds with considerable $T_\text{c}$s, Figs.~\ref{fig5}(g-j) show their isotropic superconducting gaps at various temperatures, where a larger superconducting gap indicates a stronger electron pairing interaction, requiring a higher energy to break the Cooper pairs, and is generally associated with a higher $T_\text{c}$. Table \ref{tab1} shows that the $T_\text{c}$ of the six hydrides are generally positively correlated with the $\lambda$. 

\begin{figure*}[h]
\centering
\includegraphics[width=15cm]{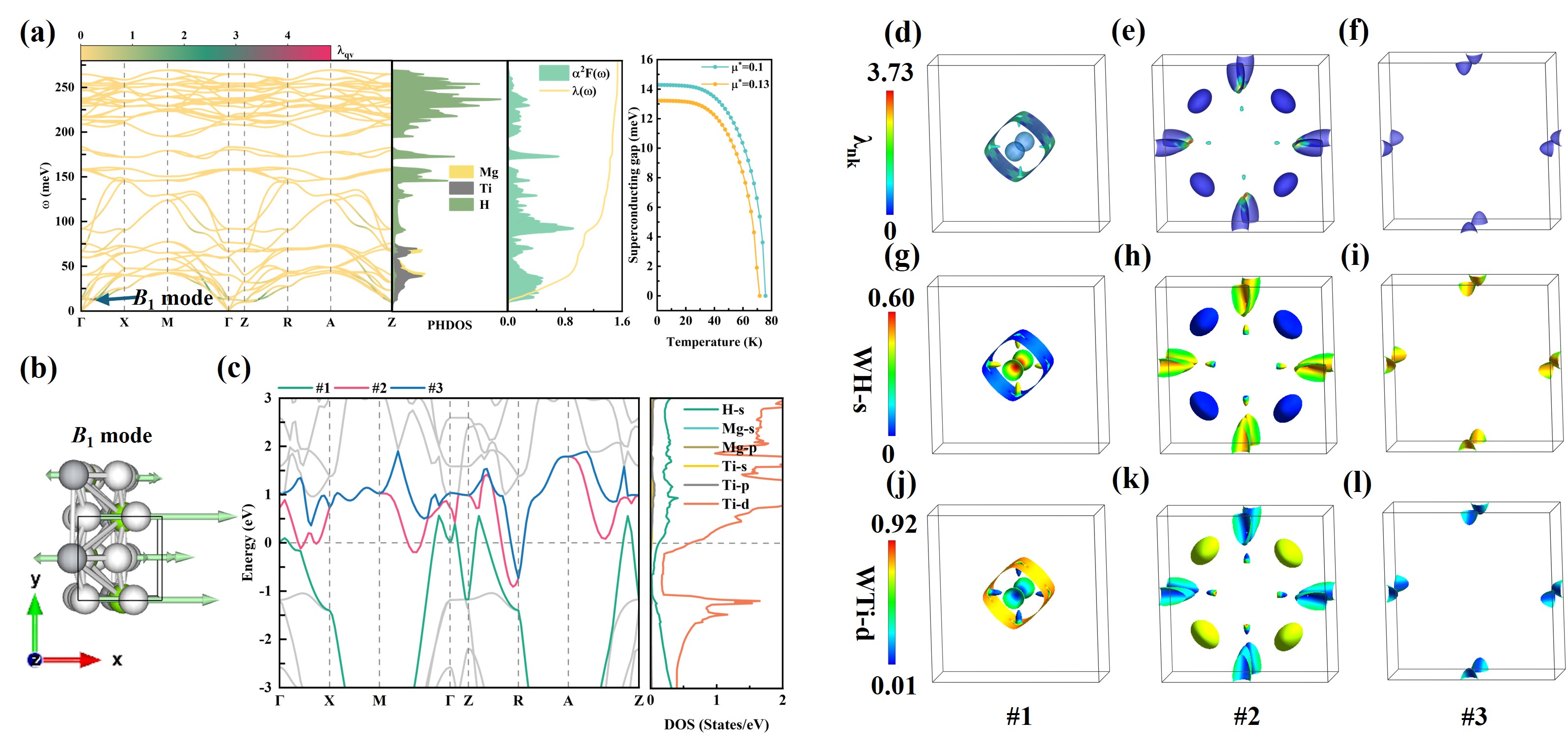}
\caption{The electronic structure, electron-phonon coupling, and superconductivity of $P4/nmm$-MgTiH$_6$ at 170 GPa. (a) Phonon dispersion curves, projected phonon density of states, $\alpha^2F(\omega)$, electron-phonon integrals $\lambda$, and isotropic superconducting gap. (b) Phonon displacements corresponding to the $B_1$ mode. (c) Electronic band structure and PDOS, with the three bands crossing the Fermi level highlighted. (d-f) Fermi surface, with colors representing the strength of EPC $\lambda_{n\bm{k}}$. (g-l) The weights of H-1$s$ and Ti-$3d$ orbitals on the Fermi surfaces. }
\label{fig6}
\end{figure*}

Up to now, all the explorations have been performed at 200 or 300 GPa, and the $P4/nmm$-MgTiH$_6$ phase exhibits the highest $T_\text{c}$ (47-62 K) among the six Mg-Ti-H compounds. In the following, we further examine whether reducing external pressure could enhance the superconducting properties of $P4/nmm$-MgTiH$_6$, potentially raising its $T_\text{c}$. Remarkably, it achieves a record-high $T_\text{c}$ of 81.9 K at 170 GPa\footnote{For the sake of computational resources, the pressures are explored with an interval of 10 GPa in this work.}, exceeding the boiling point of liquid nitrogen. As shown in Fig. \ref{fig6}(a), the $P4/nmm$-MgTiH$_6$ phase remains dynamically stable down to 170 GPa. Compared with 200~GPa, the EPC of the low-frequency phonon modes is significantly enhanced at 170 GPa, with their contribution to the $\lambda$ increasing from 55\% to 68\%, resulting in $\lambda=1.54$. Notably, the $\lambda_{q\nu}^{\rm max}$ increases to 4.83 and is associated with the $B_1$ mode of the acoustic branch along the $\Gamma$-X path, which corresponds to out-of-plane vibrations of H and metal atoms in the YZ plane [Figs.~\ref{fig6}(a, b)]. Fig.~\ref{fig6}(c) highlights the three bands crossing the Fermi level. We further map the momentum $\textbf{k}$- and band-index $n$-resolved EPC strength $\lambda_{n\bm{k}}$ on the Fermi surfaces [Figs.~\ref{fig6} (d-f)]. Figs.~\ref{fig6}(d, g, j) show that the EPC on the Fermi surface formed by band 1 is mainly associated with Ti $3d$ electrons. Figs. \ref{fig6}(e, h, k) indicate that the strongest EPC on band 2 arises predominantly from H $1s$ electrons. Figs.~\ref{fig6}(f, i, l) show that band 3 makes only negligible contributions to the EPC. The $T_\text{c}$ of $P4/nmm$-MgTiH$_6$ at 170 GPa is estimated to be in the range of 61.9-81.9~K (see also Table~\ref{tab1}), suggesting that this phase is a potential high-temperature superconductor under this pressure.

\subsection{Element substitution in the hydrogen-rich phases \texorpdfstring{$Pm\bar{3}m$-Mg$_3$TiH$_{12}$ and $P4/nmm$-MgTiH$_6$}{Pm-3m-Mg3TiH12 and P4/nmm-MgTiH6}}

In the Mg-Ti-H system, it turns out hydrides still require relatively high pressures to maintain dynamical stability, such as $P4/nmm$-MgTiH$_6$ (170 GPa) and $Pm\bar{3}m$-Mg$_3$TiH$_{12}$ (300 GPa). To reduce the required pressures and improve material performance, here we employ a simple tuning strategy: element substitution. Previous studies have shown that the introduction of heavier elements can enhance the chemical precompression effect~\cite{Zhang2022prb, Jiang2024fr, Zhao2024nsr}, thereby lowering the stability pressures of hydrides. Within this Mg-Ti-H system, $P4/nmm$-MgTiH$_6$ exhibits the highest $T_\text{c}$, whereas $Pm\bar{3}m$-Mg$_3$TiH$_{12}$ possesses the highest crystallographic symmetry, which could also be favorable with high $T_\text{c}$. Therefore, we select these two representative structures and examine how substituting Ti with heavier group-IV elements (Zr and Hf) influences their structural stability and superconducting properties.

\begin{figure}[t]
\centering
\includegraphics[width=8.5cm]{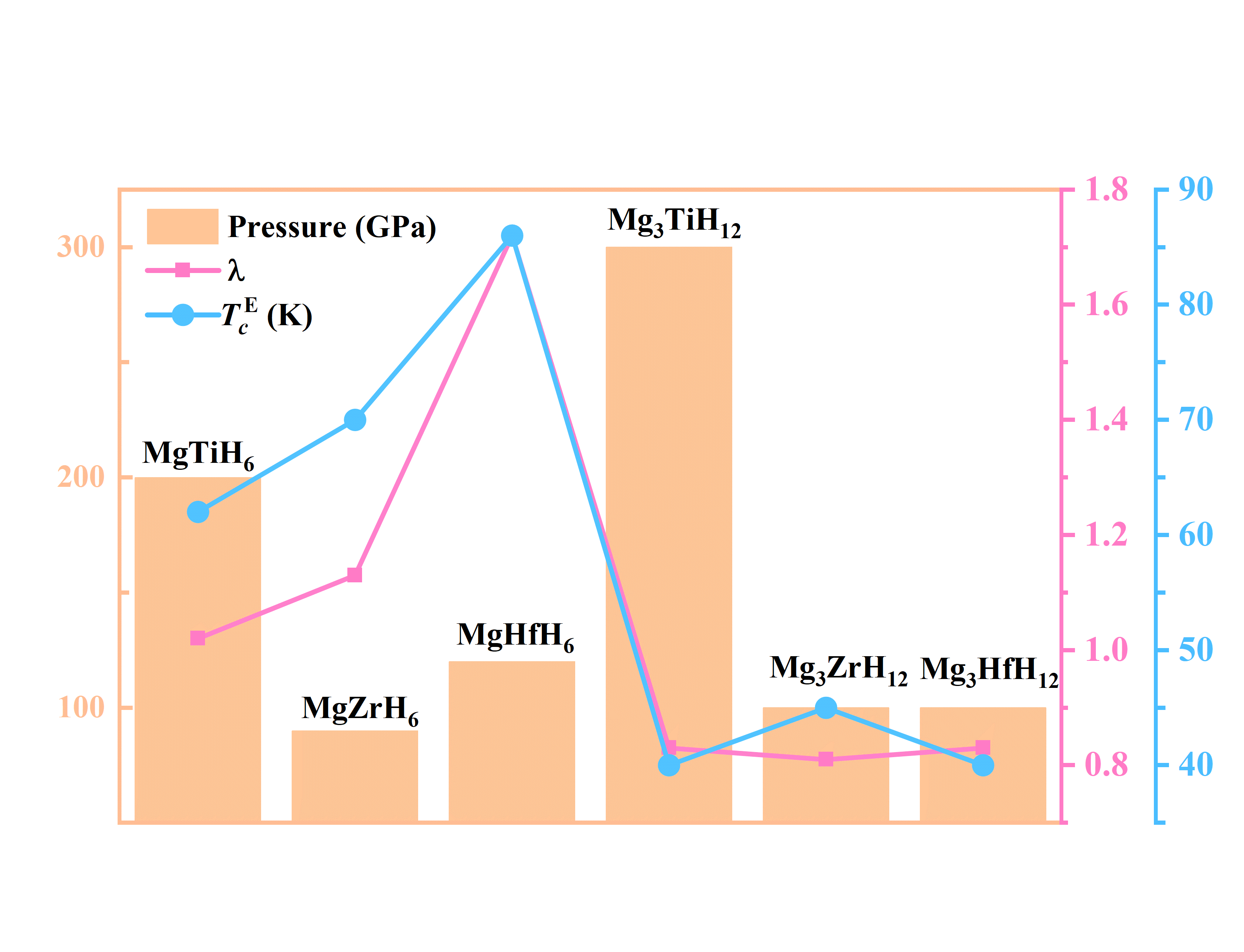}
\caption{The EPC strength $\lambda$ and $T_\text{c}$ of $P4/nmm$-MgXH$_6$ and $Pm\bar{3}m$-MgXH$_6$ (X=Ti, Zr, Hf) at their respective stability pressures. $T_\text{c}$ is evaluated using the isotropic Midgal-Eliashberg equation, with the Coulomb pseudopotential $\mu^*$ set to 0.1.}
\label{fig7}
\end{figure}

Notably, the element-substitution strategy significantly reduces the pressure required to achieve dynamical stability in both structures. As shown in Fig. S2 in the SM, the phonon dispersion of the  $P4/nmm$ and $Pm\bar{3}m$ phases after substitution exhibits no imaginary modes at 90 (120) GPa and 100 (100) GPa, respectively. For the $P4/nmm$ phase, element substitution has a pronounced impact on its superconducting properties. As shown in Fig. \ref{fig7}, both the EPC strength and the $T_\text{c}$ increase with the mass of the substituting elements. The EPC strength reaches 1.13 in $P4/nmm$-MgZrH$_6$ and 1.72 in $P4/nmm$-MgHfH$_6$. As a result, the superconducting transition temperature increases significantly, see also Table~\ref{tab1}. Notably, the predicted $T_\text{c}$ of $P4/nmm$-MgHfH$_6$ reaches 86 K, which also exceeds the boiling point of liquid nitrogen, highlighting its potential as a high-temperature superconductor. For the $Pm\bar{3}m$ phase, on the other side, Fig. \ref{fig7} reveals that although the incorporation of Zr and Hf markedly lowers the dynamical stability pressure, such a reduction of pressure is not accompanied by an obvious change in its superconducting critical temperature. The EPC strength $\lambda$ and the $T_\text{c}$ of $Pm\bar{3}m$-Mg$_3$Zr(Hf)H$_{12}$ remain essentially unchanged comparing to that of $Pm\bar{3}m$-Mg$_3$TiH$_{12}$, see also Table~\ref{tab1}. We attribute this behavior to the fact that Mg$_3$Zr(Hf)H$_{12}$ and Mg$_3$TiH$_{12}$ share highly similar electronic structures near the Fermi level (see Fig.~S3 in the SM), indicating that substituting heavier same-group elements is insufficient to induce significant electronic-structure modifications in the $Pm\bar{3}m$ phase.

\begin{figure}[th]
\centering
\includegraphics[width=8.5cm]{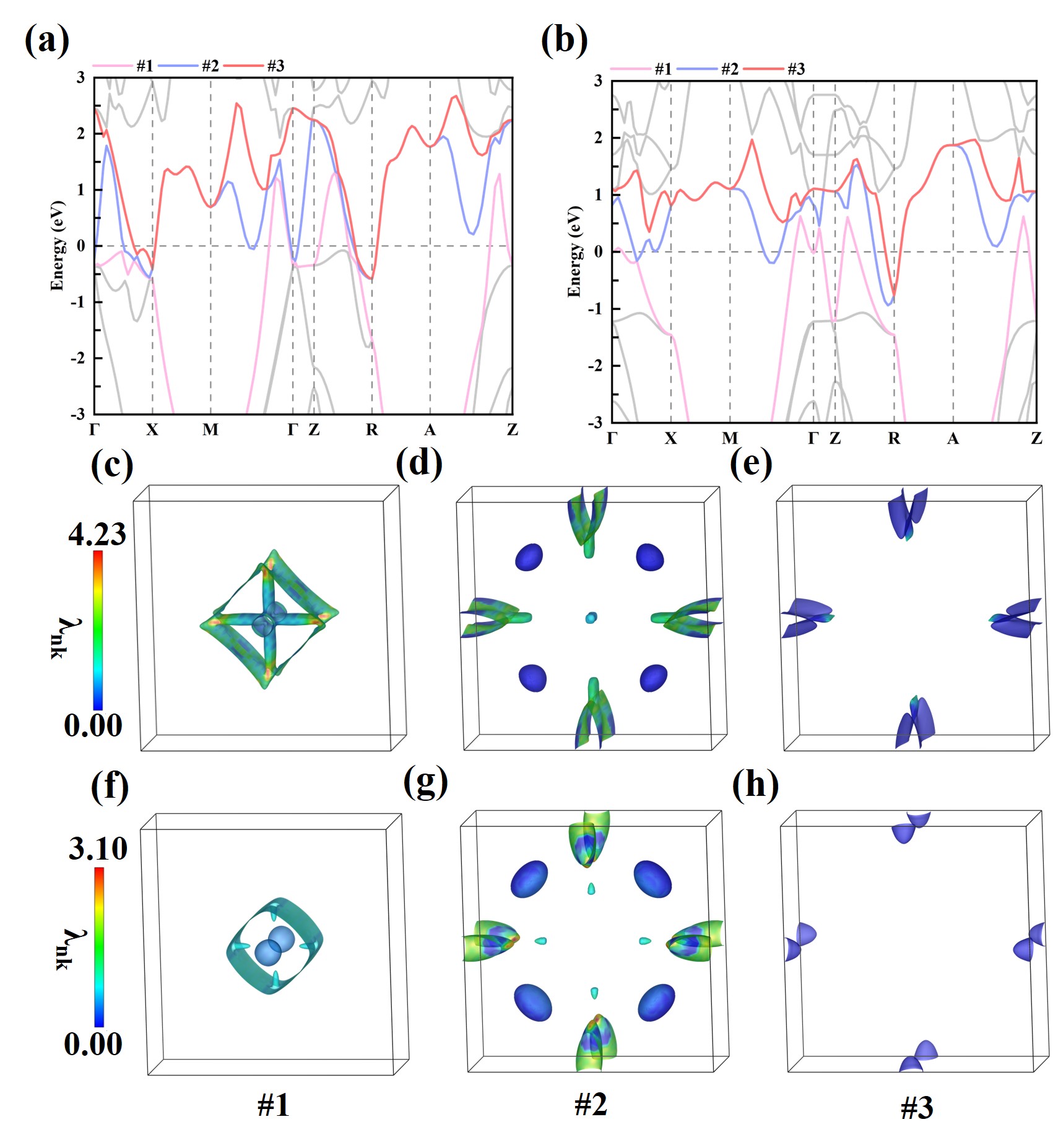}
\caption{ (a) Electronic band structure of  $P4/nmm$-MgHfH$_6$ at 120 GPa. (b) (a) Electronic band structure of $P4/nmm$-MgTiH$_6$ at 200 GPa. (c-e) Fermi surface of $P4/nmm$-MgHfH$_6$ at 120 GPa. (f-h) Fermi surface of $P4/nmm$-MgTiH$_6$ at 200 GPa. Color in (c-h) represents the EPC strength $\lambda_{n\bm{k}}$.}
\label{fig8}
\end{figure}

\begin{figure}[th]
\centering
\includegraphics[width=8.5cm]{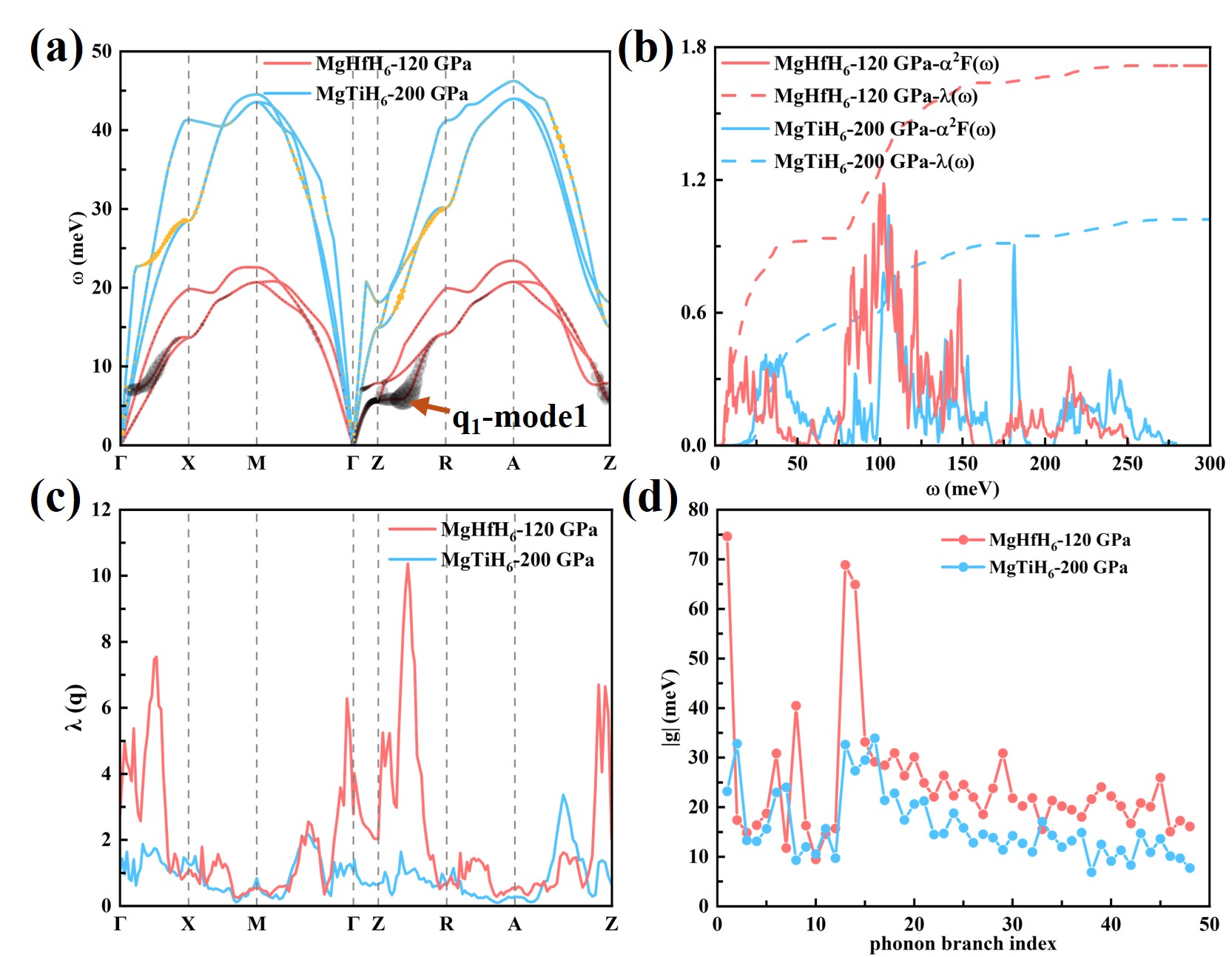}
\caption{(a) Phonon dispersion of acoustic modes of $P4/nmm$-MgHfH$_6$ at 120 GPa and $P4/nmm$-MgTiH$_6$ at 200 GPa. The size of the circle represents the EPC strength $\lambda_{\bm{q}\nu}$. (b) $\alpha^2F(\omega)$, electron-phonon integrals $\lambda$ of $P4/nmm$-MgHfH$_6$  at 120 GPa and $P4/nmm$-MgTiH$_6$ at 200 GPa. (c) The EPC strength distributions of $P4/nmm$-MgHfH$_6$ and $P4/nmm$-MgTiH$_6$ along high symmetry paths. (d) Average EPC matrix element $|\bm{g}|$ of various phonon branches at $\bm{q}_1$-point.}
\label{fig9}
\end{figure}

In the following, we examine in detail the origins of the enhancement in superconducting properties observed in $P4/nmm$-MgHfH$_6$, comparing with $P4/nmm$-MgTiH$_6$. As shown in Figs. \ref{fig8}(a-b), the electronic structure of MgHfH$_6$ near the Fermi level exhibits pronounced changes relative to MgTiH$_6$, which is expected to affect the EPC in this phase. Figs. \ref{fig8}(c-h) further illustrate the differences in the Fermi surfaces between the two phases, revealing an overall enhancement of the EPC on the MgHfH$_6$ Fermi surface compared to that of MgTiH$_6$. This enhancement is primarily associated with the low-frequency acoustic branches of MgHfH$_6$ [see Fig. S2(d) in the SM]. For clarity, Fig.~\ref{fig9}(a) shows the acoustic branches of MgHfH$_6$ and MgTiH$_6$ for comparison. The acoustic branches of MgHfH$_6$ are significantly softened, and the phonon softening along the $\Gamma$-X and Z-R paths substantially enhances the EPC strength. As shown in Fig. \ref{fig9}(b), the spectral function $\alpha^2F(\omega)$ peaks in the acoustic branch region of MgHfH$_6$ clearly shift to lower frequencies relative to that of MgTiH$_6$, in good correspondence with Fig. \ref{fig9}(a). Fig.~\ref{fig9}(c) shows the variations of the EPC strength along high-symmetry paths. The peaks of EPC strength in MgHfH$_6$ coincide with the $\bm{q}$ points where phonon softening occurs. The most significant difference relative to MgTiH$_6$ is observed at the $\bm{q}_1$ point marked in Fig. \ref{fig9}(a). Further analysis in Fig. \ref{fig9}(d) reveals that the average EPC matrix elements $|\bm{g}| $ at this $\bm{q}_1$ point are enhanced compared to those of MgTiH$_6$. Notably, the first phonon mode at the $\bm{q}_1$ point (with the lowest frequency, marked as $\bm{q}_1$-mode1 in Fig. \ref{fig9}(a)) exhibits the most significant enhancement of the average EPC matrix element, indicating that this softened phonon mode significantly strengthens electron scattering. The corresponding vibrational pattern of the $\bm{q}_1$-mode1 is shown in Fig. S4 in the SM.

\section{Conclusions}

In summary, various structures of Mg$_x$TiH$_{2y}$ ($x=1$-3, $y=3$-8) were predicted under high pressures of 200~GPa and 300~GPa. Two new thermodynamically stable phases, $P4/nmm$-MgTiH$_6$ and $Pmm2$-Mg$_3$TiH$_6$, were identified at both 200~GPa and 300 GPa. Thermodynamically stable Mg$_3$TiH$_{12}$ was also identified, whose space group is $R3/m$ at 200~GPa and $Pm\bar{3}m$ at 300 GPa, respectively. The superconductivity of these four stable structures together with two metastable phases, $I4_1amd$-MgTiH$_8$ and $P4/nmm$-MgTiH$_{10}$, had been investigated in detail. Among these Mg-Ti-H structures, $P4/nmm$-MgTiH$_6$ exhibits the best superconducting properties: it achieves a record-high $T_\text{c}$ of 81.9~K at 170 GPa, exceeding the boiling point of liquid nitrogen. Such a high $T_\text{c}$ is primarily attributed to strong EPC driven by low-frequency acoustic phonon modes. In addition, the heavy-element substitution strategy successfully reduced the pressures required for dynamical stability of the hydrogen-rich, high-symmetric phases, i.e., the $Pm\bar{3}m$ and $P4/nmm$ phases. The dynamical stability pressures of $Pm\bar{3}m$-Mg$_3$ZrH$_{12}$ and $Pm\bar{3}m$-Mg$_3$HfH$_{12}$ were lowered to 100 GPa. For the $P4/nmm$ phase, MgZrH$_6$ and MgHfH$_6$ remain dynamically stable down to 90 GPa and 120 GPa, respectively. In $P4/nmm$-MgHfH$_6$, the modification of the electronic structure near the Fermi level and the softening of low-frequency phonon modes both contribute to a significant enhancement of the EPC strength. As a result, its $T_\text{c}$ increases to 86 K, making it a second candidate whose $T_\text{c}$ exceeds the boiling point of liquid nitrogen. Our study provides valuable insights for future experiments and can serve as a useful guide for realizing potential high-temperature superconductors in such ternary hydrides.

\section*{CRediT authorship contribution statement}

\textbf{Min Pan}: Resources, Writing - Review \& Editing, Supervision, Project administration.
\textbf{Yujie Wang}: Methodology, Validation, Formal analysis, Investigation, Data Curation, Writing - Original Draft, Writing - Review \& Editing, Visualization.
\textbf{Kaige Hu}: Software, Writing - Review \& Editing, Supervision.
\textbf{Huiqiu Deng}: Resources, Writing - Review \& Editing

\section*{Declaration of Competing Interest}

The authors declare that they have no known competing financial interests or personal relationships that could have influenced the work reported in this paper.

\section*{Data availability}

Data will be made available on request.

\section*{Acknowledgments}

This work is financially supported by the Sichuan Science and Technology Program (Grant No. 2026YFHZ0030) and the National Natural Science Foundation of China (No. 12375259). We gratefully acknowledge HZWTECH for providing computational facilities.

\appendix

\section{Supplementary Material}

Supplementary Material to this article can be found online at \url{doi.org/10.1016/j.mtp.2026.xxxxxx}. 

\bibliographystyle{elsarticle-num} 
\bibliography{References}

\newpage

\begin{center}
\large
\emph{Supplementary Material} for \\ 
First-principles study on the high-$T_\text{c}$ superconductivity of Mg-Ti-H ternary hydrides up to the liquid-nitrogen temperature range under high pressures
\\
\bigskip
\normalsize
\centerline{Min Pan$^\text{a}$, Yujie Wang$^\text{a, b}$, Kaige Hu$^\text{c,*}$, Huiqiu Deng$^\text{d}$}
\smallskip
\small
$^a$\textit{School of Electrical Engineering, Southwest Jiaotong University, Chengdu, 610031, China} \\
$^b$\textit{School of Materials Science and Engineering, Southwest Jiaotong University, Chengdu, 610031, China} \\
$^c$\textit{School of Physics and Optoelectronic Engineering \& Guangdong Provincial Key Laboratory of Sensing Physics and System Integration Applications, Guangdong University of Technology, Guangzhou, 510006, China}\\
$^d$\textit{School of Physics and Electronics, Hunan University, Changsha, 410082, China}\\
$^*$Corresponding author. \textit{Email address:} \texttt{hukaige@gdut.edu.cn} (Kaige Hu)
\end{center}

\setcounter{equation}{0}
\setcounter{figure}{0} 
\setcounter{page}{1}
\setcounter{table}{0} 
\renewcommand{\theequation}{S\arabic{equation}}
\renewcommand{\thefigure}{S\arabic{figure}}
\renewcommand{\thepage}{S\arabic{page}}
\renewcommand{\thetable}{S\arabic{table}}

\begin{figure*}[h]
\centering
\includegraphics[width=15cm]{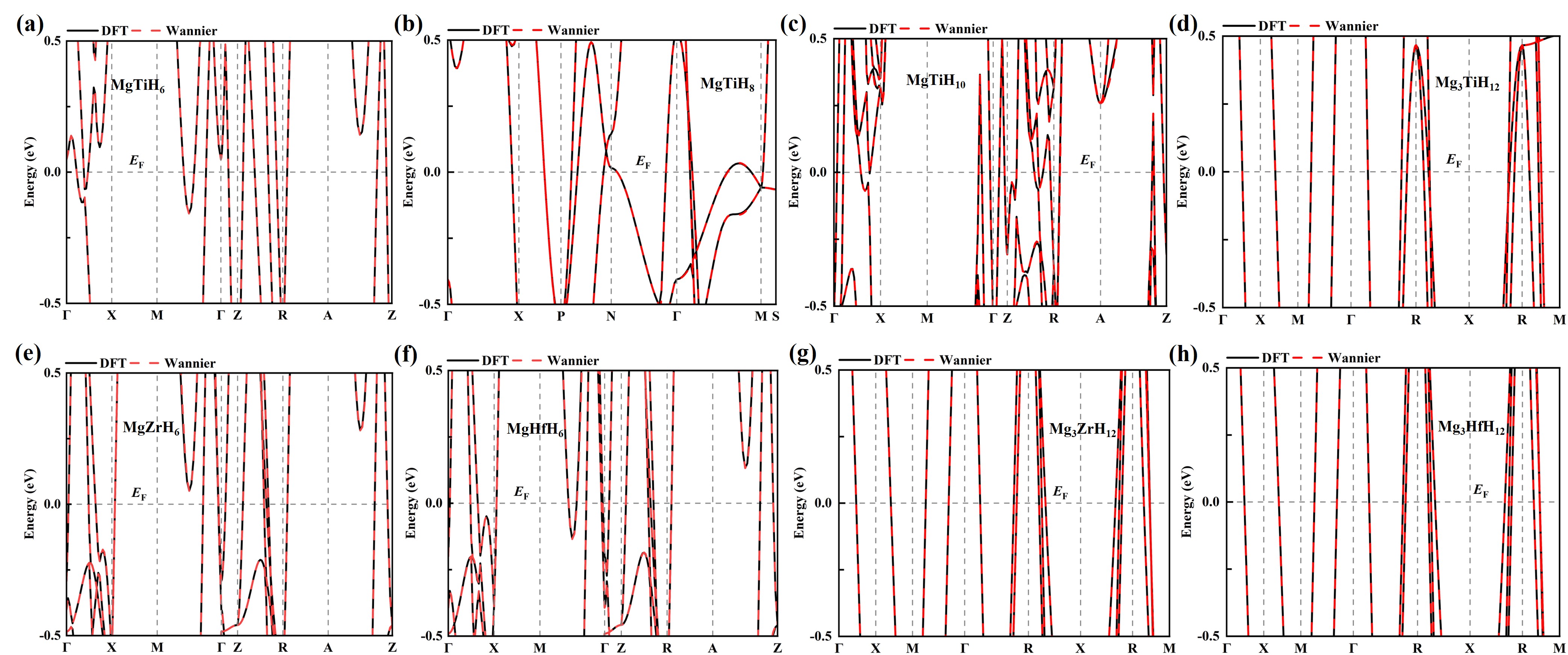}
\caption{(a-h) The electronic band structure near the Fermi level for eight compounds. The black solid lines correspond to DFT calculations, while the red dashed lines indicate Wannier interpolation. The Fermi level is set to zero.}
\label{figS1}
\end{figure*}

\begin{figure*}[h]
\centering
\includegraphics[width=12cm]{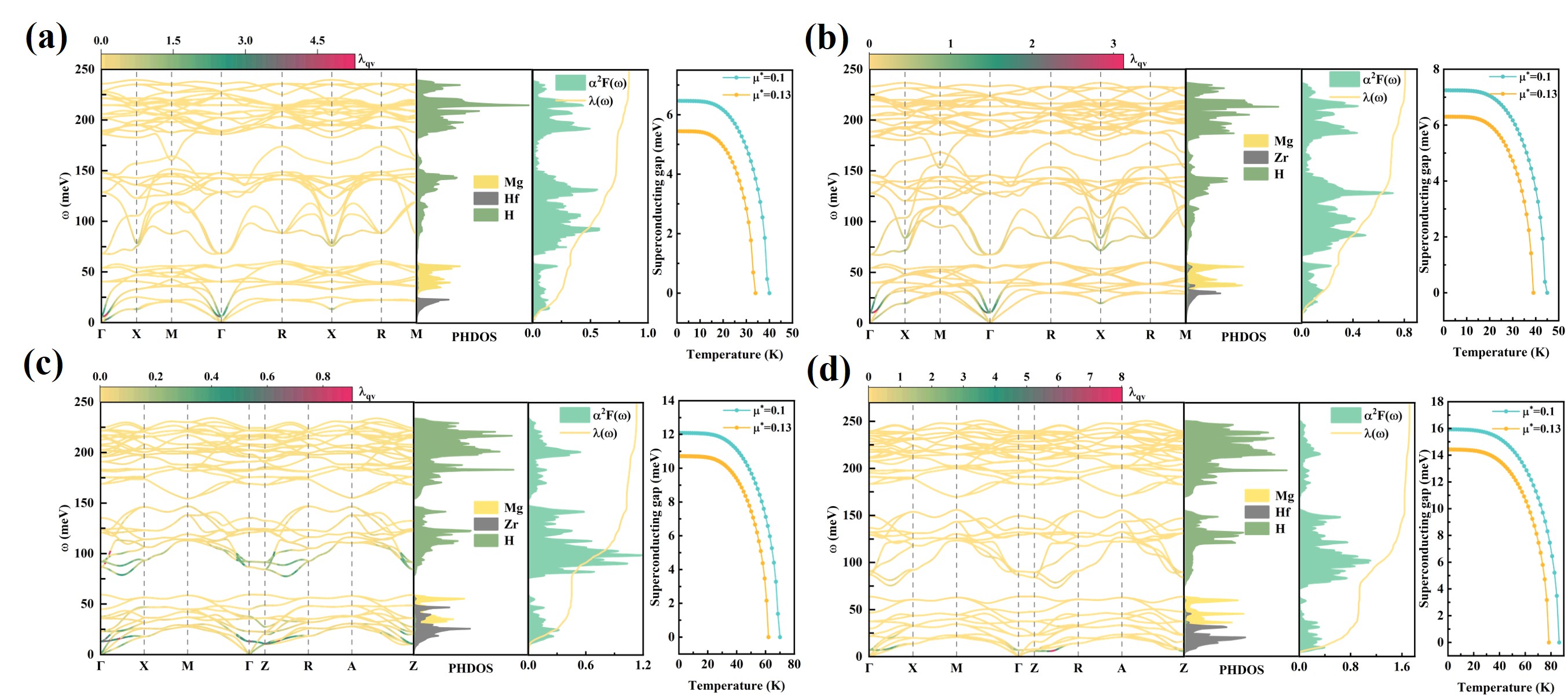}
\caption{Phonon dispersion curves, projected phonon density of states, $\alpha^2F(\omega)$, electron-phonon integrals $\lambda$ and isotropic superconducting gap of (a) $Pm\bar{3}m$-Mg$_3$ZrH$_{12}$ at 100 GPa, (b) $Pm\bar{3}m$-Mg$_3$HfH$_{12}$ at 100 GPa, (c) $P4/nmm$-MgZrH$_6$ at 90 GPa, and (d) $P4/nmm$-MgHfH$_6$ at 120 GPa.
Phonon dispersion curves of (a) $Pm\bar{3}m$-Mg$_3$HfH$_{12}$ at 200 GPa, (b) $Pm\bar{3}m$-Mg$_3$ZrH$_{12}$ at 200 GPa, and (c) $Pm\bar{3}m$-Mg$_3$HfH$_{12}$ at 100 GPa.}
\label{figS2}
\end{figure*}

\begin{figure*}[th]
\centering
\includegraphics[width=8.5cm]{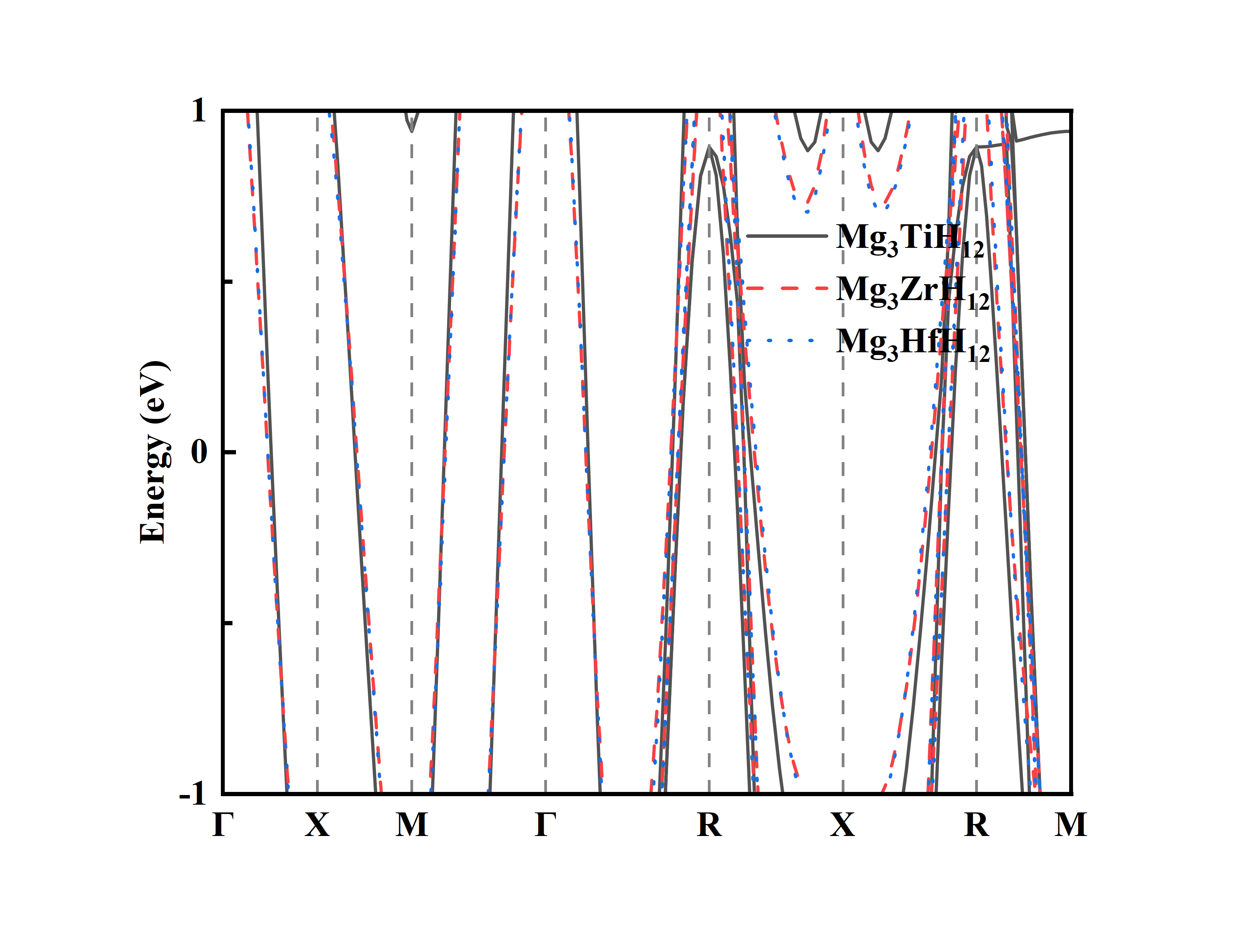}
\caption{Electronic band structure of $Pm\bar{3}m$-Mg$_3$TiH$_{12}$, $Pm\bar{3}m$-Mg$_3$ZrH$_{12}$ and $Pm\bar{3}m$-Mg$_3$HfH$_{12}$.}
\label{figS3}
\end{figure*}

\begin{figure}[th]
\centering
\includegraphics[width=7.5cm]{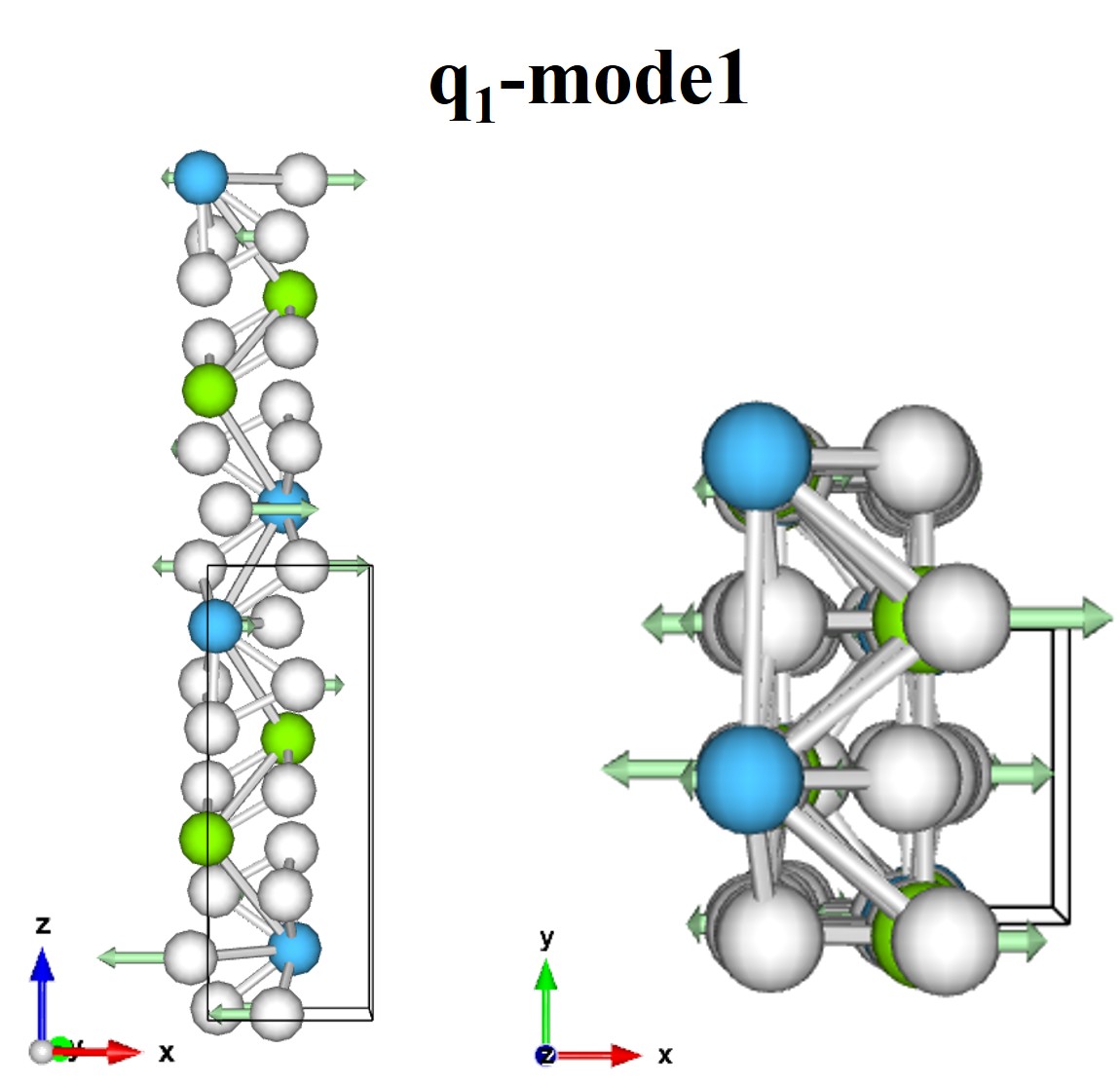}
\caption{The vibrational pattern of the $\bm{q}_1$-mode1 phonon.}
\label{figS4}
\end{figure}

\begin{table*}[th]
\centering
\caption{Structural parameters of various Mg-Ti-H compounds}\label{tabS1}
\begin{tabular}{ccccrc}
\toprule[1pt]
Compound & \makecell{Space \\ group} & \makecell{Lattice \\ 
parameters}  & Atoms & \makecell{Atomic coordinates \\ (fractional) \\ $x$ \qquad $y$ \qquad $z$} & \makecell{Wyckoff \\ position} \\ 
\midrule[1pt]
\makecell{MgTiH$_6$ \\ (200 GPa)} & $P4/nmm$ & \makecell{$a=b=2.682$ \AA \\ $c=7.293$ \AA \\ $\alpha=\beta=\gamma=90^\circ$ \\} & \makecell{Mg \\ Ti \\ H1 \\ H2 \\ H3 \\ H4 \\ H5} & \makecell{0.0000 0.5000 0.3811 \\ 0.0000 0.5000 0.8740 \\ 0.0000 0.5000 0.1209 \\ 0.0000 0.0000 0.2487 \\ 0.0000 0.0000 0.5000 \\ 0.5000 0.0000 0.3523 \\ 0.5000 0.5000 0.0000} & \makecell{2$c$ \\ 2$c$ \\ 2$c$ \\ 4$f$ \\ 2$b$ \\ 2$c$ \\ 2$a$} \\ \midrule
\makecell{Mg$_3$TiH$_6$ \\ (200 GPa)} & $Pmm2$ & \makecell{$a=2.634$ \AA \\ $b=3.656$ \AA \\ $c=4.481$ \AA \\ $\alpha=\beta=\gamma=90^\circ$ \\} & \makecell{Mg1 \\ Mg2 \\ Mg3 \\ Ti \\ H1 \\ H2 \\ H3 \\ H4} & \makecell{0.0000 0.0000 0.4457 \\ 0.5000   0.0000 0.9186 \\ 0.0000 0.5000 0.1199 \\ 0.5000 0.5000 0.5937 \\ 0.0000 0.0000 0.0965 \\ 0.5000 0.5000 0.9380 \\ 0.0000 0.2716 0.7498 \\ 0.5000 0.2482 0.2810} & \makecell{1$a$ \\ 1$c$ \\ 1$b$ \\ 1$d$ \\ 1$a$ \\ 1$d$ \\ 2$g$ \\ 2$h$} \\ \midrule
\makecell{Mg$_3$TiH$_{12}$ \\ (200 GPa)} & $R3m$ & \makecell{$a=b=5.265$ \AA \\ $c=6.451$ \AA \\ $\alpha=\beta=90^\circ$ \\ $\gamma=120^\circ$} & \makecell{Mg \\ Ti \\ H1 \\ H2 \\ H3 \\ H4 \\ H5 \\ H6} & \makecell{$-0.1634$ $-0.1634$ $-0.7145$ \\ 0.0000 0.0000 $-0.3710$ \\ 0.0000   0.0000 $-0.8950$ \\ 0.0000 0.0000 $-0.6235$ \\ 0.0000 0.0000 $-0.1219$ \\ 0.1690 0.3381 $-0.2994$ \\ $-0.1643$ $-0.3286$ $-0.4659$ \\ $-0.3667$ $-0.1833$ $-0.5517$} & \makecell{9$b$ \\ 3$a$ \\ 3$a$ \\ 3$a$ \\ 3$a$ \\ 9$b$ \\ 9$b$ \\ 9$b$} \\ \midrule
\makecell{MgTiH$_8$ \\ (300 GPa)} & $I4_1amd$ & \makecell{$a=b=4.413$ \AA \\ $c=5.262$ \AA \\ $\alpha=\beta=\gamma=90^\circ$} & \makecell{Mg \\ Ti \\ H} & \makecell{0.0000 0.0000 0.5000 \\ 0.0000 $-0.5000$ 0.2500 \\ $-0.1548$  0.1479 0.2474} & \makecell{4$b$ \\ 4$a$ \\ 32$i$} \\ \midrule
\makecell{MgTiH$_{10}$ \\ (300 GPa)} & $P4/nmm$ & \makecell{$a=b=2.583$ \AA \\ $c=8.579$ \AA \\ $\alpha=\beta=\gamma=90^\circ$} & \makecell{Mg \\ Ti \\ H1 \\ H2 \\ H3 \\ H4 \\ H5 \\ H6} & \makecell{0.0000 0.5000 0.0959 \\ 0.5000 0.0000 0.3058 \\ 0.0000   0.5000 0.2959 \\ 0.5000 0.0000 0.1153 \\ 0.5000 0.0000 0.4942 \\ 0.0000 0.0000 0.0000 \\ 0.8293 0.0000 0.5784 \\ 0.5000 0.5000 0.2022} & \makecell{2$c$ \\ 2$c$ \\ 2$c$ \\ 2$c$ \\ 2$c$ \\ 2$a$ \\ 8$i$ \\ 4$f$} \\ \midrule
\makecell{Mg$_3$TiH$_{12}$ \\ (300 GPa)} & $Pm\bar{3}m$ & \makecell{$a=b=c=3.565$ \AA \\ $\alpha=\beta=\gamma=90^\circ$} & \makecell{Mg \\ Ti \\ H1 \\ H2 \\ H3} & \makecell{0.0000 0.5000 0.0000 \\ 0.0000 0.0000 0.0000 \\ 0.5000 0.5000 0.5000 \\ 0.0000 0.5000 0.0000 \\ 0.7473 0.2527 0.7473} & \makecell{3$c$ \\ 1$a$ \\ 1$b$ \\ $3d$ \\ 8$g$} \\
\bottomrule[1pt]
\end{tabular}
\end{table*}

\end{document}